\documentclass[aps,pre,twocolumn,groupedaddress,showpacs,superscriptaddress,amssymb,amsmath]{revtex4-1}
\usepackage{amsmath}
\usepackage{latexsym}
\usepackage{amssymb}
\usepackage{graphics,epstopdf}
\usepackage{hyperref}
\usepackage{xcolor}
\hypersetup{
    colorlinks = true,
    linkcolor =blue,
	citecolor=blue, 
	urlcolor=blue 
}

\newcommand{\ep}{\epsilon}
\newcommand{\la}{\langle}
\newcommand{\ra}{\rangle}

\newcommand{\ed}{\end{document}}
\newcommand{\beq}{\begin{equation}}
\newcommand{\eeq}{\end{equation}}

\usepackage{mathtools}
\usepackage{comment}

\begin{document}
\title{Quantum transport in quasi-periodic lattice systems in presence of B{\"u}ttiker probes}
\author{Madhumita Saha}
\email{madhumita.saha@acads.iiserpune.ac.in }
\affiliation{Department of Physics, Indian Institute of Science Education and Research Pune, Dr. Homi Bhabha Road, Ward No. 8, NCL Colony, Pashan, Pune, Maharashtra 411008, India}
\author{B. Prasanna Venkatesh}
\email{prasanna.b@iitgn.ac.in}
\affiliation{Department of Physics, Indian 
Institute of Technology Gandhinagar, Palaj, Gujarat 382355, India}

\author{Bijay Kumar Agarwalla}
\email{bijay@iiserpune.ac.in}
\affiliation{Department of Physics, Indian Institute of Science Education and Research Pune, Dr. Homi Bhabha Road, Ward No. 8, NCL Colony, Pashan, Pune, Maharashtra 411008, India}

\begin{abstract}
Quasi-periodic lattice systems offer diverse transport properties. In this work, we investigate the environment induced effects on transport properties for quasi-periodic systems, namely the one-dimensional Aubry-Andr\'e-Harper (AAH) lattice chain and its generalized version (GAAH) by considering the B{\"u}ttiker probe approach. We first consider voltage probe situation and study the electrical conductance properties in the linear response regime. At zero temperature, we observe enhancement in conductance for small probe coupling strength $\gamma$ with a power-law scaling $\gamma^4$ at all the no-transport regimes, located both inside and outside of the band of the original system. Whereas, for large probe coupling strengths, the conductance at all Fermi energies is the same and decays as a power-law with scaling $1/\gamma^4$. This particular scaling survives even in the finite-temperature limit. Interestingly, this scaling  result is different from the one recently predicted following the local Lindblad master equation approach. The transport eventually becomes diffusive for all energy ranges and in all regimes of the original model for a sufficiently strong coupling with the probes. We further extend our study and consider voltage-temperature probes to analyze the thermoelectric performance of the chain in terms of the figure of merit. We also demonstrate the validity of two recently obtained bounds on thermoelectric efficiency which are tighter than the seminal Carnot bound, in presence of voltage-temperature probes.
\end{abstract}
\date{\today}
\maketitle

\section{Introduction}
Quasi-periodic lattice systems reside between a completely disordered system and a periodically ordered system. Such systems can offer a lot of interesting and intriguing transport properties even in low dimensions \citep{Aubry2, Harper, Commensurate_AAH,GAAH_mobility_edge,Pandit83, Sumilan, Modak}. Recent remarkable experimental realizations of quasi periodic systems in various cold atom platforms \cite{I_bloch_experiment, zilberberg1,zilberberg2,GAAH_experiment, expt1,expt2,expt3,AAH-expt-Silberberg,mobility_edge_expt}  have triggered intense theoretical research activities to understand the underlying dynamical and steady-state transport properties. Moreover, some new studies in this context have further revealed the potential of realising highly efficient quantum devices such as quantum rectifiers, thermoelectric engines by carefully exploiting the exotic transport properties \cite{device-1, device-2, quasi-periodic_rectifier, quasi-periodic-rectifier3}. 

The popular quasi-periodic lattice systems, namely, the Aubry-Andr\'e-Harper (AAH), its generalised version GAAH model and the Fibonacci model \cite{Fibonacci-first,Fibonacci-second} have been studied extensively in the context of boundary driven dissipative quantum transport \citep{Archak_phase_diagram, Archak_AAH, Madhumita_GAAH, Madhumita_readout,Varma_quasi_periodic,interaction_quasi_periodic,interaction_Fibonacci}.  Further studies have started to emerge to understand the environment induced effects on transport  \cite{dephasing_disorder,dephasing-interaction1, dephasing_spin_interaction2, dephasing_interaction3, Landi-latest,ENQT1,buttikerdephase,dephasing_dubi} in such systems. Very recently, following the local Lindblad master equation formalism, the steady-state transport properties due to dephasing noise were analyzed for AAH and Fibonacci models \cite{Landi-latest}. 
Following a similar approach, the effect of dephasing noise on transport was studied in presence of mobility edge \cite{ENQT1}. 
Apart from this approach, another popular way to incorporate environment effects is the B{\"u}ttiker probe technique. Such a technique is used extensively to understand effective many-body transport properties in setups like molecular junctions \cite{molecular_dephasing1, molecular_dephasing2, molecular_dephasing3, molecular_dephasing4, segal-probe-1,segal-probe-2,segal-probe-3,malay}, quantum dots \cite{Bedkihal2013}, lattice models of oscillators, spins, fermions, excitons \cite{Pastawski, Roy-1,self-2-Abhishek, dephasing_dubi} etc. 
Very recently authors in Ref. \cite{buttikerdephase} extended the B{\"u}ttiker probe approach to analyze thermoelectric transport properties in Fibonacci type quasi-periodic models \cite{Bu-1,Bu-2,Bu-3,Bu-4}  and pointed out that environment induced processes can lead to a better thermoelectric performance in certain regimes of transport. However a clear understanding of transport and thermoelectric properties for other types of quasi-periodic systems such as  the AAH and the GAAH model (quasi-periodic system with a single-particle mobility edge) following the B{\"u}ttiker probe approach is still missing.

In this work, we therefore analyse how transport properties in AAH and GAAH  lattice models get modified in presence of the B\"uttiker probes. We implement both the voltage probe and voltage-temperature probe techniques and explore electrical conductance and thermoelectric heat-to-work conversion properties in the linear-response regime. We point out the consequences for the conductance in both zero and finite temperature limits. We observe that in presence of both the voltage and the voltage-temperature probes, the linear transport coefficients, namely the electrical and the thermal conductances display universal power-law decay with the probe coupling strength in the strong probe coupling regime. We provide a rigorous analytical proof for this result. Importantly, this universal behaviour is different from the behaviour predicted following the local Lindblad master equation formalism, thus pin-pointing crucial differences between the two approaches in mimicking the underlying many-body scattering processes. Furthermore, in the zero temperature voltage-probe case, we observe environment assisted transport for all the exponentially decaying regimes with a particular power-law scaling with probe strength. For the thermoelectric setup, we further asses and compare the recently obtained tighter bounds on efficiency following the thermodynamic uncertainty relations \cite{Saryal-bound} and the bound obtained in Ref. \cite{Udo-TUR1}.

We organize the paper as follows: In section II, we introduce the AAH and the GAAH lattice models and then briefly describe the theoretical aspects of the B{\"u}ttiker probe approach and how to obtain charge conductance in the linear response regime. In section III, we present the numerical results both in the zero and finite temperature limit. In section IV, we extend our study to voltage-temperature probes and study thermoelectric performance for the AAH chain and furthermore asses the recently obtained tighter bounds on efficiency by expressing these bounds in terms of the Onsager's transport coefficients. Finally in section V, we provide a summary highlighting our key findings. In the Appendix, we present a rigorous derivation and show the emergence of diffusive behaviour for conductance as well as the its scaling with the probe coupling strength in the strong probe coupling limit. 

\section{Model and theory}
\subsection{Model Hamiltonian}
We consider an out-of-equilibrium one-dimensional quasi-periodic lattice chain that is connected at its two ends to two thermo-chemical reservoirs. These reservoirs are always maintained at fixed chemical potentials and temperatures. Since we are interested in investigating the environment induced effects on the central quasi-periodic
lattice chain, we connect uniform and independent local reservoirs, i.e., B\"uttiker probes to each lattice site. The Hamiltonian for the entire setup is then given as 
\begin{align}
H&=H_S\!+\! H_B\!+\! H_{SB}\!+\! H_{P} \!+\! H_{PS},
\label{Eq.1}
\end{align}
where $H_S$ represents a quasi-periodic chain with Hamiltonian 
\begin{align}
\label{equation2}
H_S=\sum_i \epsilon_i c_i^{\dagger}c_i + t \sum_i \big(c_{i+1}^{\dagger} c_i +c_{i}^{\dagger} c_{i+1}\big).
\end{align}
Here $c_{i }^{\dagger} (c_{i})$ is the creation (annihilation) operator for the electron at the $i$th site. Here we consider only the nearest neighbour hopping with  hopping amplitude $t$. The site dependent onsite potential $\epsilon_i$  mimics a quasi-periodic potential which for the generalized Aubry-Andr{\'e}-Harper (GAAH) model is given as,
\begin{align}
\epsilon_i= \frac{2\lambda \cos[2\pi b i + \Phi]}{1+\alpha \cos[2 \pi b i + \Phi]},
\label{onsite}
\end{align}
where $\lambda$ represents the strength of the potential, $b$ is an irrational number which makes the potential quasi-periodic, and $\Phi$ is the phase factor that generates different configurations of the quasi-periodic potential. For, $\alpha=0$, this model reduces to the AAH model \cite{Harper, Aubry2}. For  AAH model, all the single particle eigenstates are delocalized for $\lambda<1$ and exponentially localized for $\lambda>1$ \cite{Aubry2} and at $\lambda=1$ all the states are critical, i.e., neither localized nor delocalized. As a result, the particle transport goes from ballistic to exponentially suppressed as the value of $\lambda$ is tuned from delocalized to localized regime and at the critical point the transport shows anomalous behaviour with sub-diffusive scaling of transport coefficient with system size \cite{Archak_AAH,Madhumita_GAAH}. 

For non-zero $\alpha$, i.e., for the GAAH model, the system possesses a single-particle mobility edge which can be obtained analytically,
$E= 2 \,{\rm sgn}{(\lambda)(1-|\lambda|)}/\alpha$ \cite{GAAH_mobility_edge} where $E$ corresponds to a single particle energy eigenvalue. In this case, all the single particle eigenstates with energy less than $E$ are extended/delocalized while those states with energy higher than $E$ are localized. Hence, when $E$ falls within the spectrum, it is a mobility edge.

In Eq.~(\ref{Eq.1}),  $H_B=H_L + H_R$ represents the Hamiltonian for the left and the right reservoirs. For this study, we model these reservoirs as $1$-D tight-binding ordered semi-infinite chain with Hamiltonian,
\begin{align}
\label{leftbath}
H_L&=\epsilon_0 \sum_k  a_k^{\dagger} a_k + t_0 \sum_k  a_{k+1}^{\dagger} a_k + {\rm H.c}, \\ 
H_R&=\epsilon_0 \sum_k  b_k^{\dagger} b_k + t_0 \sum_i b_{k+1}^{\dagger}  b_k +{\rm H.c}, 
\end{align}
with $a_k$ and $b_k$ correspond to the annihilation operators of the $k$th sites for left and right baths, respectively. For simplicity, we choose the same onsite ($\epsilon_0)$ and the same hopping parameter ($t_0$) for both the baths. The term $H_{SB}= H_{SL} + H_{SR}$ represents the coupling between the system and the baths and is responsible for the charge transfer across the system. It is specified by,
\begin{align}
\label{couplingH}
H_{SL} &=\gamma_L a_1^{\dagger} c_1 + {\rm H.c}, \\ 
H_{SR}&=\gamma_R b_1^{\dagger} c_N + {\rm H.c},  
\end{align}
where the left (right) bath is coupled to the first ($N$-th) site of the quasi-periodic lattice chain with coupling strength $\gamma_L$ ($\gamma_R$). 

\begin{figure}
\includegraphics[trim=5 38 5 40, clip, width=\columnwidth]{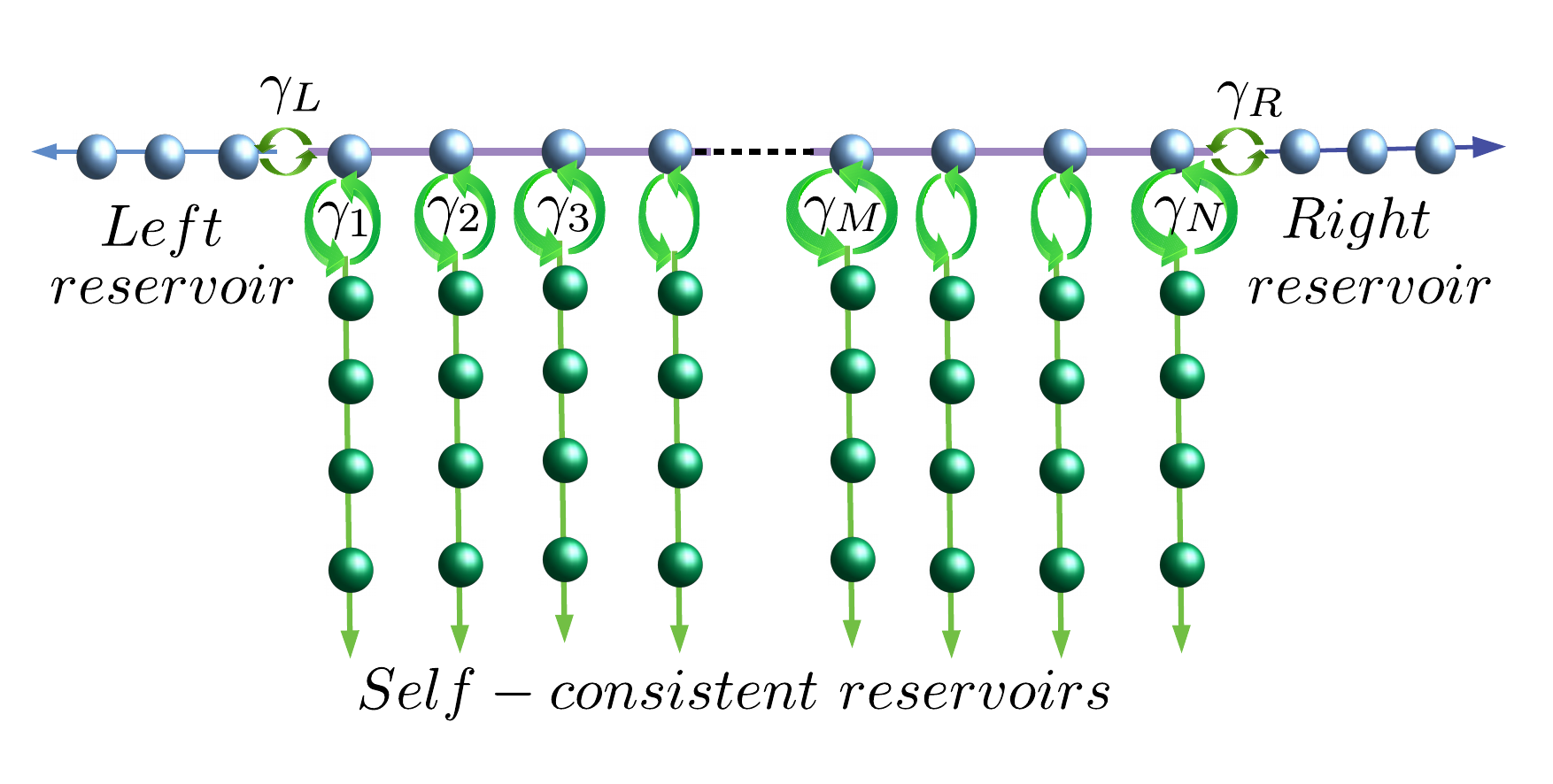} 
\caption{(Color online): Schematic of a one-dimensional lattice chain with the first and the last sites of the chain are connected to the left and the right reservoirs, respectively, These reservoirs have different chemical potentials that induce current in the chain. In addition, each lattice site is attached to a local reservoir which we refer here as the B\"uttiker  probe. The probes, left and right reservoirs are all modelled as 1-D ordered semi-infinite tight binding chains.}
\label{schematic}
\end{figure}
As mentioned before, the dephasing and dissipation effects within the central lattice chain can be modelled phenomenologically by connecting local reservoirs (probes) at each lattice site. As done for the left and the right reservoirs, for this study, we model each probe also as a 1D tight-binding ordered semi-infinite chain. The total Hamiltonian for the probes is given as $H_P = \sum_{i=1}^{N} H_P^{i}$ with each probe Hamiltonian is given by
\begin{align}
H_P^{i}= \sum_j \, \Big[\epsilon_0 d_{ij}^{\dagger} d_{ij}+ t_0 d_{ij+1}^{\dagger} d_{ij}+ {\rm  H.c} \Big].
\end{align}
Correspondingly, the coupling Hamiltonian between the $i$-th probe and the central system is given as,
\begin{align}
H^{i}_{PS}=\gamma c_i^{\dagger} d_{i1} + {\rm H.c},
\end{align}
where the first site of each probe is coupled to the $i$-th site of the central system with coupling strength $\gamma$, which is chosen to be the same for all the probes (see Fig.~(\ref{schematic})).
\subsection{Theory}
We employ the non-equilibrium Green's function (NEGF) approach \cite{Keldysh1,Keldysh2,Keldysh3,Keldysh4,Wang2014,Wingreen} to investigate end-to-end transport through the central quasi-periodic chain. In presence of the two-baths at the ends and the B{\"u}ttiker probes, one can obtain the retarded Green's function for the central system as,
\begin{align}
{\cal G}^r(\ep)=\Big[\ep I \!-\! H_C \!-\! \Sigma^r_L(\ep)\!-\!\Sigma^r_R(\ep) \!-\!\sum_{i=1}^N \Sigma^r_{P,i}(\ep)\Big]^{-1}
\end{align}
where $I$ is the $N \times N$ Identity matrix with $N$ being the number of lattice sites of the central system and is also equal to the total number of probes attached to the central lattice. $H_C$ represents $N\times N$ single particle hamiltonian corresponding to $H_S$. $\Sigma^r_{\alpha}(\ep), \alpha=L, R, P$ is the self-energy associated with the $\alpha$-th bath. Note that the effects of the baths, including the probes, are additive in the self-energy. Since we have chosen the baths and the B\"uttiker probes as semi-infinite ordered tight-binding chains, the self-energies in this case can be obtained analytically and given by \cite{bound-NEGF},
\begin{align}
\Sigma^r_{\alpha}(\ep)=\frac{\gamma_{\alpha}^2}{2\, t^{2}_0}\Big[\ep \!-\!i \sqrt{4 t_0^2 \!-\! \ep^2}\Big], \quad \alpha=L, R, P.
\end{align}
where $\epsilon_0=0$, and $\gamma_P = \gamma$. Before we proceed further, let us fix some notations. We identify the index $n$ with the probe terminals, $\nu$ to identify the left and the right reservoirs and use index $\alpha$ to count all the leads. Since the entire setup is fully quadratic, the average charge current in the steady state flowing out of the $\nu$-th reservoir is given by the famous Landauer-Buttiker formula \cite{Keldysh4}, 
\begin{align}
I_{\nu}=\frac{e}{2\pi} \sum_{\alpha} \int_{-\infty}^{\infty} d\ep \, \mathcal{T}_{\nu \alpha}(\ep) \, \Big(f_{\nu} (\ep)-f_{\alpha}(\ep)\Big), \quad \nu=L,R.
\end{align}
Here $f_{\alpha}(\ep)=(1+e^{\beta(\ep-\mu_{\alpha})})^{-1}$ is the Fermi distribution function of $\alpha$-th terminal with inverse temperature $\beta$ and chemical potential $\mu_{\alpha}$. $\mathcal{T}_{\nu \alpha}(\ep)$ is the transmission probability for an electron to flow from $\nu$-th terminal to $\alpha$-th terminal through the quasi periodic system.
The transmission probabilities can be computed
using the Green’s function of the central system and the
self-energy of the baths and expressed as,
\begin{align}
\mathcal{T}_{\nu \alpha}(\ep)={\rm Tr} \Big [\Gamma_{\nu}(\ep)\,{\cal G}^r(\ep)\,\Gamma_{\alpha} (\ep)\, {\cal G}^a (\ep)\Big],
\end{align}
where ${\cal G}^a(\epsilon)=\big[{\cal G}^a(\epsilon)\big]^{\dagger} $ is the advanced Green's function and $\Gamma_{\alpha}(\ep)=-2\, {\rm Im} \big[\Sigma^r_{\alpha}(\ep)\big]$.

One can similarly compute the charge current flowing out of the $n$th probe which is also given by the Landauer-Buttiker formula,
\begin{align}
\label{probe_current}
I_{n}=\frac{e}{2\pi} \sum_{\alpha} \int_{-\infty}^{\infty} \, d\ep \, \mathcal{T}_{n \alpha}(\ep) \Big(f_{n} (\ep)-f_{\alpha}(\ep)\Big), \, \, n=1,2, \cdots N.
\end{align}
In what follows,  we first analyze the effect on end-to-end conductance properties by considering the local reservoirs as voltage probes and thus set the net charge current flowing out of each probe to zero, i.e., $I_{n} =0$ for $n=1, 2, \cdots N$ in Eq.~(\ref{probe_current}). Imposing this constraint for each probe allows one to simulate both elastic dephasing and inelastic dissipative processes in the central lattice chain.  In section IV, we further extend the above study by considering the local reservoirs as voltage-temperature probes and demand vanishing charge and heat currents from each probe which then allow us to simulate inelastic but non-dissipative scattering processes. Under this generic probe setting, we study thermoelectric heat-to-work conversion properties and the recently obtained tighter bounds on thermoelectric efficiency.

\subsection{Voltage probe technique in linear response regime}
In this sub-section, we first focus on the voltage probe technique by considering the temperature of the end reservoirs and the probes to be the same. The  different chemical potential of the two end reservoirs drive a steady-state current in the chain. We impose the zero charge current condition from each probe which in turn fixes the local chemical potential of each probe. We concentrate on the linear response regime and expand the Fermi-distribution function around the equilibrium chemical potential $\ep_F$ and inverse temperature $\beta$ as, 
\begin{align}
\label{linear_response}
f_{K} (\ep)=f_{\rm eq}(\ep)-\frac{\partial f_{\rm eq}(\ep)}{\partial \ep} (\mu_{K} - \ep_F).
\end{align}
The derivative of the Fermi function is also evaluated at $\ep_F$. Now putting Eq.~(\ref{linear_response}) in Eq~(\ref{probe_current}), for each $n$ and setting $I_n = 0, n=1, 2, \cdots N$, we receive a set of $N$ linear equations for the chemical potential of each probe $n$ of the form,
\begin{eqnarray}
\label{probesolution}
\mu_n&=& \mu_R \!+\! \sum_{j=1}^N \Big[{\cal W}_{nj}^{-1} \int_{-\infty}^{\infty} \mathcal{T}_{jL}(\ep) (-\frac{\partial f_{\rm eq} (\ep)}{\partial \ep}) d\ep \Big] (\mu_L \!-\! \mu_R), \nonumber \\
&&\,\, \forall \,\,\,n=1, 2, \cdots N 
\end{eqnarray}
Here ${\cal W}$ is an $N\times N$ matrix  given as,
\begin{widetext}
${\cal W}=\begin{pmatrix}
\sum_{\alpha \neq 1} \int_{-\infty}^{\infty} \mathcal{T}_{1\alpha}(\ep) \Big(-\frac{\partial f_{\rm eq} (\ep)}{\partial \ep}\Big) d\ep & -\int_{-\infty}^{\infty} \mathcal{T}_{12}(\ep) (-\frac{\partial f_{\rm eq} (\ep}{\partial \ep}) d\ep & -\int_{-\infty}^{\infty} \mathcal{T}_{13}(\ep) (-\frac{\partial f_{\rm eq} (\ep)}{\partial \ep}) d\ep  & \ldots \\
-\int_{-\infty}^{\infty} \mathcal{T}_{21}(\ep) \Big(-\frac{\partial f_{\rm eq} (\ep)}{\partial \ep}\Big) d\ep & \sum_{\alpha \neq 2} \int_{-\infty}^{\infty} \mathcal{T}_{2\alpha}(\ep) (-\frac{\partial f_{\rm eq} (\ep)}{\partial \ep}) d\ep & -\int_{-\infty}^{\infty} \mathcal{T}_{23}(\ep) (-\frac{\partial f_{\rm  eq} (\ep)}{\partial \ep}) d\ep & \ldots \\ 
-\int_{-\infty}^{\infty} \mathcal{T}_{31}(\ep) (-\frac{\partial f_{\rm eq} (\omega)}{\partial \ep}) d\omega & -\int_{-\infty}^{\infty} \mathcal{T}_{32}(\ep) (-\frac{\partial f_{\rm eq} (\ep)}{\partial \ep}) d\ep & \sum_{\alpha \neq 3} \int_{-\infty}^{\infty} \mathcal{T}_{3\alpha}(\ep) (-\frac{\partial f_{\rm eq} (\ep)}{\partial \ep}) d\ep & \ldots \\
\ldots & \ldots & \ldots & \ldots 
\end{pmatrix}$.
\end{widetext}
Given the local chemical potentials $\mu_n$ in Eq.~(\ref{probesolution}), we can compute the electrical conductance $G$  for this setup as,
$G={I_R}/\Delta V$ with $\Delta V={(\mu_R \!-\! \mu_L)}/e$ being the applied bias voltage,
\begin{align}
G &= G_0\, \int_{-\infty}^{\infty} \, d\ep  \Big(-\frac{\partial f_{\rm eq} (\ep)}{\partial \ep}\Big)   \Big[\mathcal{T}_{RL}(\ep)+ \nonumber \\
& \int_{-\infty}^{\infty} \, d\ep' \sum_{n j}  \mathcal{T}_{R n}(\ep) \,\, {\cal W}^{-1}_{nj} \,\, \mathcal{T}_{jL}(\ep') \Big(-\frac{\partial f_{\rm eq} (\ep')}{\partial \ep'}\Big)\Big]
\end{align}
where $G_0 = e^2 / 2 \pi \hbar$  is the universal quantum electrical conductance. The second term in the above expression reflects the change in the conductance due to the probes by mimicking incoherent scattering processes. 
In the zero temperature limit ($\beta \to \infty$) \cite{Pastawski}, the above equation simplifies drastically as $\frac{\partial f_{\rm eq} (\ep)}{\partial \ep}=-\delta(\ep-\ep_F)$, and the conductance can be expressed simply in terms of an effective transmission function $G=G_0 \, \mathcal{T}_{\rm eff} (\ep_F)$, where 
\begin{equation}
\mathcal{T}_{\rm eff} (\ep_F)=\mathcal{T}_{R L}(\ep_F)+\sum_{n j} \mathcal{T}_{R n}(\ep_F) \,\, \mathcal{W}^{-1}_{nj,0} \,\, \mathcal{T}_{jL}(\ep_F)
\label{eff-trans}
\end{equation}
 In the above equation, $ \mathcal{W}_{nj,0}$ denotes the matrix elements of ${\cal W}$ computed in the zero temperature limit.
Interestingly, for $\beta \to \infty$, the local probes are exactly equivalent to the dephasing probes as the zero particle current condition from Eq.~(\ref{probe_current}) is now satisfied for each energy. 
 
			
			 \begin{figure*}
\includegraphics[width=2.0\columnwidth]{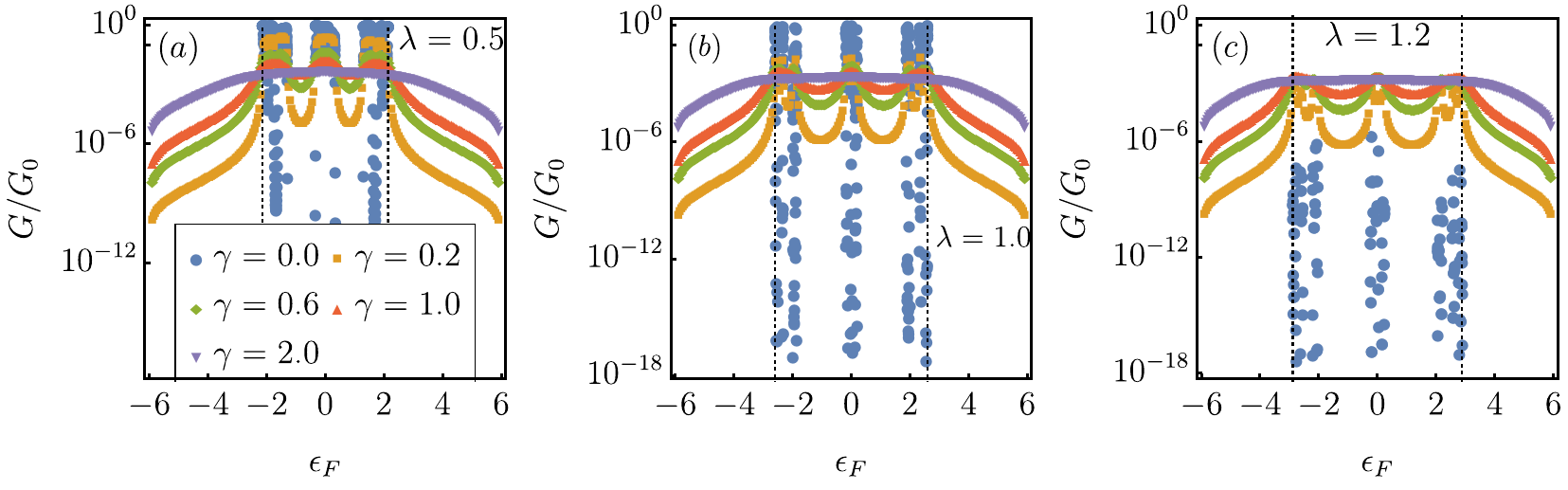} 
\caption{(Color online): Plots for zero-temperature conductance in absence and presence of B\"uttiker voltage probes for the AAH model in three different transport regimes. In the coherent regime (absence of probes), AAH model displays delocalized ($\lambda <1$), critical ($\lambda =1$), and localized ($\lambda >1$) phases depending on the strength $\lambda$. Here, we display the effect of voltage probes on transport. The vertical dotted lines represent the band edges of the AAH model. Here, the system size is taken as $N=200$.}
\label{fig1}
\end{figure*}

\begin{figure*}
\includegraphics[width=2.0\columnwidth]{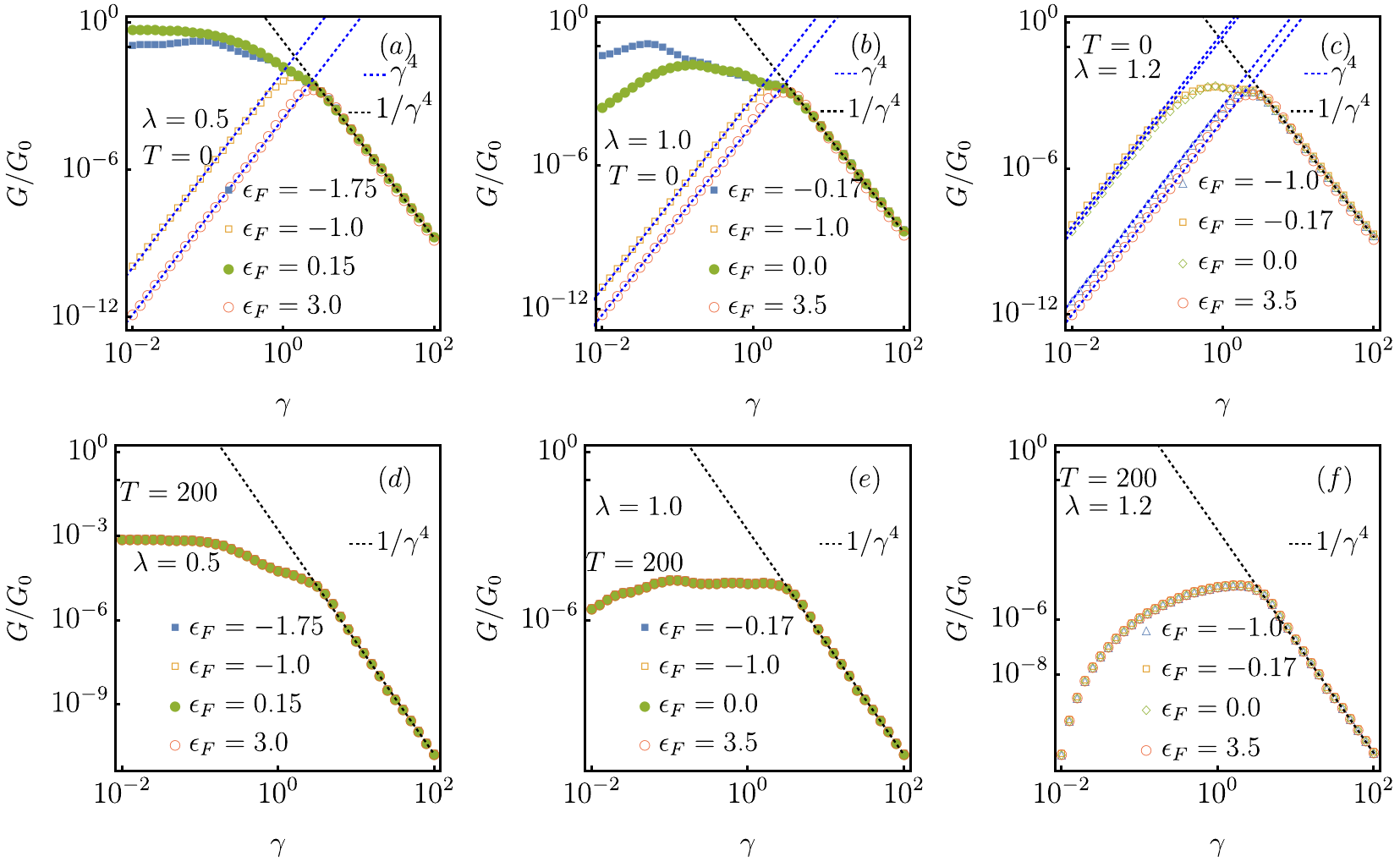} 
\caption{(Color online):  Plots for zero-temperature conductance as a function of system-probe coupling strength $\gamma$ in different transport regimes of AAH model at different Fermi energies.  Here, the system size is taken as $N=200$.}
\label{fig2} 
\end{figure*}

\begin{figure}
\includegraphics[width=\columnwidth]{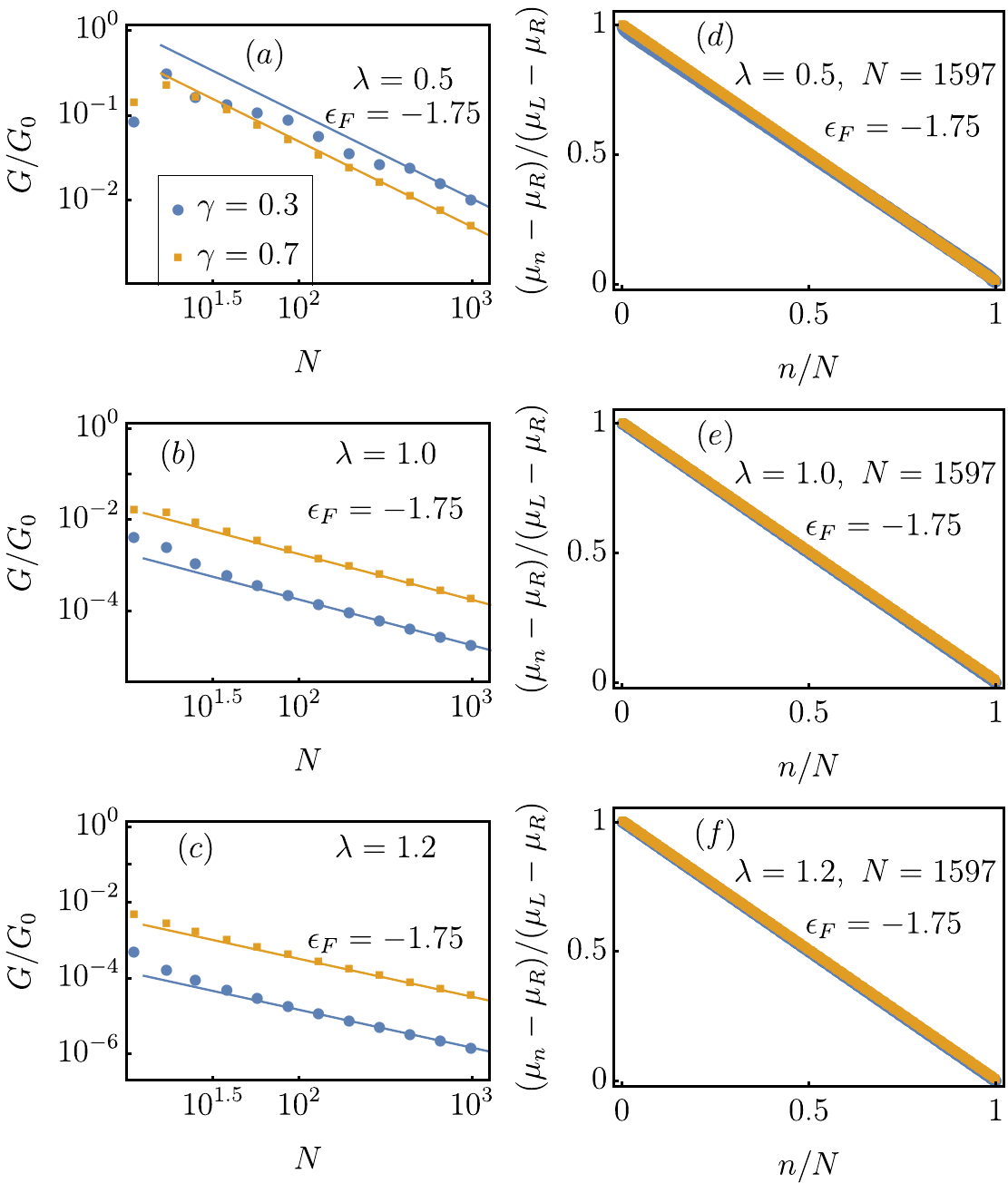} 
\caption{(Color online): Plots (a)-(c) show crossover to diffusive transport regime from ballistic ($\lambda=0.5$), sub-diffusive ($\lambda=1.0$) and exponentially localized ($\lambda=1.2$) phases, respectively, in presence of voltage probes at zero temperature. The solid lines represent diffusive $1/N$ scaling. Plots (d)-(f) show the corresponding chemical potential profiles. The linear chemical potential profile ensures that the transport is diffusive. We show results for two different $\gamma$ values.}
\label{fig5} 
\end{figure}

\begin{figure}
\includegraphics[width=\columnwidth]{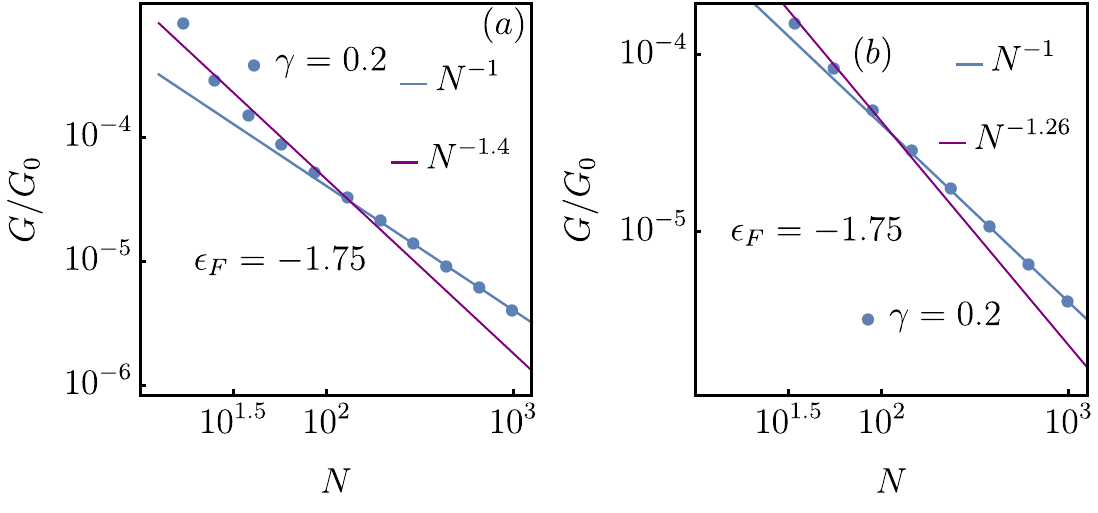} 
\caption{(Color online): Crossover from sub-diffusive behaviour to diffusive behaviour for $\gamma=0.2$ at the critical point ($\lambda=1$) of AAH model at zero temperature.  Depending on the chosen system size one can see crossover to $N^{-1}$ from different sub-diffusive scaling. In (a), the chosen system sizes do not correspond to Fibonacci numbers and in (b), the system sizes correspond to Fibonacci numbers.}
\label{figx} 
\end{figure}

\begin{figure*}
\includegraphics[width=2.0\columnwidth]{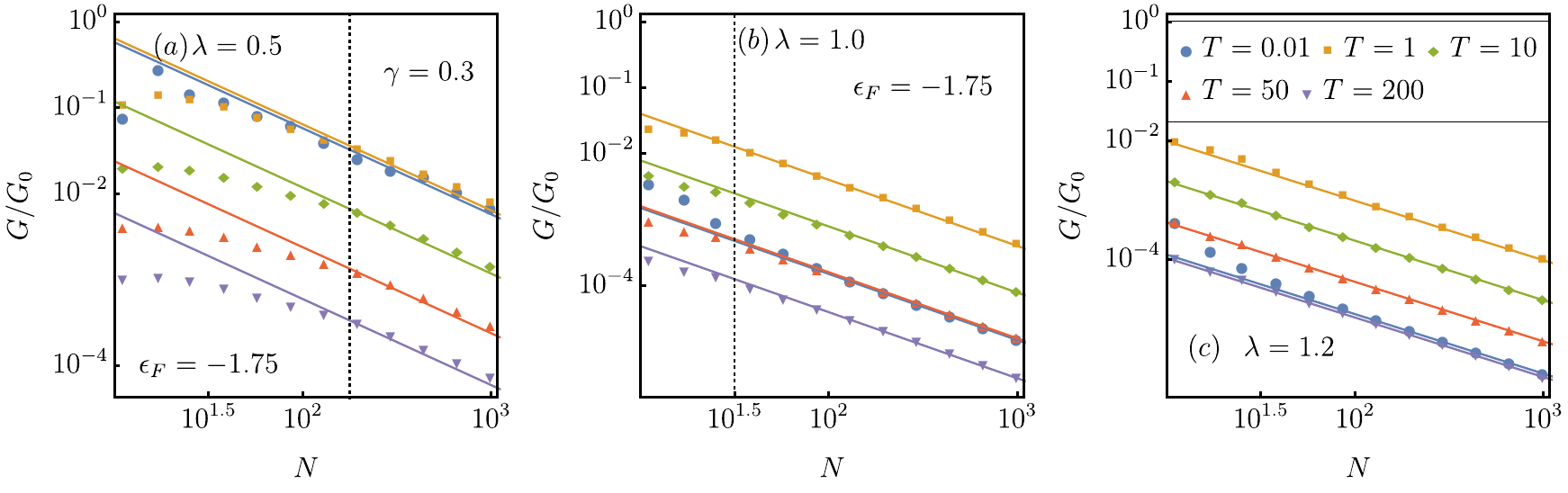} 
\caption{(Color online) Plots (a)-(c) shows crossover to diffusive regime at finite temperatures for probe coupling strength $\gamma=0.3$. The solid lines represent diffusive $1/N$ scaling. The vertical dotted lines represent the system size after which the diffusive transport sets further indicating temperature independent system-size crossover.}
\label{fig5a} 
\end{figure*}

\begin{figure}
\includegraphics[width=0.7\columnwidth]{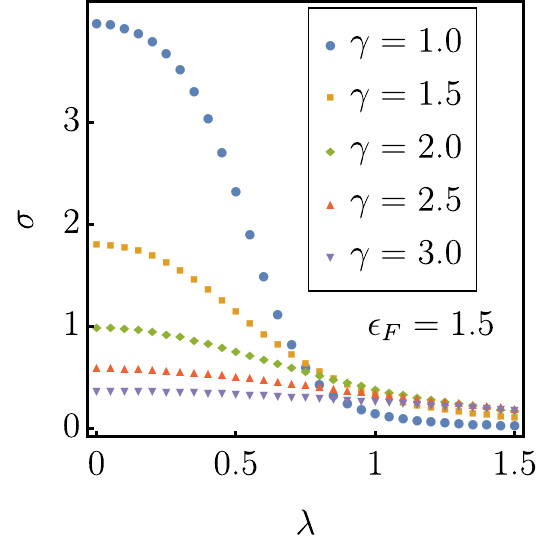} 
\caption{(Color online): Plot for electrical conductivity $\sigma= \lim_{N \to \infty} N G$ in the diffusive regime as a function of quasi-periodic lattice strength $\lambda$ for different values of $\gamma$. } 
\label{fig6} 
\end{figure}

\section{Results- voltage probe}
We now present  the numerical results for conductance at both zero and finite temperatures. Since the zero temperature calculation does not require any numerical integration to be performed, one can therefore simulate large system sizes. Here we report results upto $N=1597$.
Unless otherwise stated, for all the numerical calculations we set $\epsilon_0=0$, $t_0=3$, $t=1$, $\gamma_L=\gamma_R=1$. Note that, to reduce the number of parameters in the problem, we have chosen a spatially uniform situation for the probes by setting $\gamma_n= \gamma$ for all $n=1, 2, \cdots N$. We also stress that in the following discussions by coherent transport we imply transport through quasi-periodic lattice in absence of the probes ($\gamma_n=0$).  

In Fig.~\ref{fig1} we first display the effects of voltage probes in three different transport regimes of the AAH model corresponding to three different $\lambda$ values: ballistic ($\lambda=0.5$, Fig.~(\ref{fig1}(a))), critical ($\lambda=1.0$, Fig.~(\ref{fig1}(b))) and localized ($\lambda=1.2$, Fig.~(\ref{fig1}(c))) at zero temperature. 
In absence of probes, in all three regimes of the AAH model, there are energy values located within and outside the band of the lattice system (the vertical dotted lines in Fig.~(\ref{fig1})  represent the band edges of the central lattice chain), at which no significant transport takes place due to the absence of  bare energy states of the central system. 
However, once the voltage probes are attached, the transport properties in all the regimes change significantly. In particular, in presence of probes, finite transport is induced for all Fermi energies that corresponds to no-transport in the coherent case. Moreover, we observe that  for these particular energies, the value of conductance increases with increasing system-probe coupling $\gamma$. In contrast, the energy values for which significant transport is already present in the coherent limit, the conductance display non-monotonic behavior with increasing $\gamma$. Note that, in the localized regime (Fig.~(\ref {fig1})(c)), as there is essentially no transport for any $\epsilon_F$ in the coherent limit, attaching probes always induces finite conductance for all Fermi energies. 
 
To establish this behaviour more explicitly, we choose different $\epsilon_F$'s corresponding to no-transport and transport cases in the coherent limit, and plot in Fig.~\ref{fig2}(a)-(c) the zero-temperature conductance as a function of the coupling strength  $\gamma$. As mentioned before, for no-transport energies, the conductance increases with $\gamma$ but interestingly as a power-law with scaling $\gamma^4$ upto a critical value $\gamma_c \sim t, \gamma_L, \gamma_R$. In contrast, no particular scaling is observed for the transport energies where attaching probes may increase or decrease the value of conductance. Note that, for the critical $\lambda$ (Fig.~\ref{fig2}(b)), as the eigenstates of the  bare system are not completely delocalized, increase in conductance is observed even for Fermi energies that corresponds to finite transport.  Beyond the critical value of the probe coupling, i.e., $\gamma_c > t, \gamma_L, \gamma_R$,  the conductance becomes independent of $\lambda$ and is equal for all $\epsilon_F$'s and more importantly decays as a power law $1/\gamma^4$. Note that, this scaling result is different than the one recently predicated following the local Lindblad master equation approach where the analysis was carried out in the infinite temperature limit with scaling given as $1/\gamma^2$ \cite{Landi-latest}.
In Fig.~(\ref {fig2})(d)-(f)) we extend these results to finite-temperature limit. We observed that, at high temperature $T=200$, the conductance value is the same for all  Fermi energies, implying a flat transmission spectrum with $\epsilon_F$. Below $\gamma_c$, the contributions to conductance come from all energies that reside within the energy window of width $k_B T$ around $\epsilon_F$ and as a result no particular scaling with $\gamma$ is observed.  However, interestingly, for large $\gamma$ the conductance scaling remains the same as the zero temperature case. In Appendix A, we provide a detailed proof of this universal scaling for conductance in the strong $\gamma$ limit.

We next present the scaling of the conductance with the system size $N$. In Fig.~\ref{fig5}(a)-(c), we  display the crossover from various transport regimes in the coherent limit to a diffusive limit $G \sim N^{-1}$. We note that for finite system size, with sufficiently strong incoherent effects induced by the probes, the transport eventually becomes diffusive for all $\lambda$ values.  For weak $\gamma$, the conductance however display a crossover from the coherent limit scaling to a diffusive scaling and the corresponding crossover length varies strongly depending on whether the transport is ballistic, sub-diffusive or localized in the coherent regime. As an example, for the same $\gamma$ value ($\gamma=0.3$) in the localized case (\ref{fig5}(c)), the crossover takes place at a small system size $N \sim 30$ compared to the ballistic case $N \sim 200$.  This fact corroborates with the predictions in Ref.~(\cite{Landi-latest}).  For a fixed $\lambda$, with increasing  $\gamma$, the crossover length shrinks and a diffusive scaling sets in. In Fig.~\ref{fig5}(d)-(f), we display the corresponding chemical potential profile in the lattice chain which shows a clear linear behavior in presence of probes, indicating a diffusive transport.

It is important to note that, in the critical regime of the coherent AAH model $(\gamma=0)$, the conductance can display different sub-diffusive scalings depending on the choice of the system size \cite{Archak_AAH, Varma_quasi_periodic}. For example, at $\lambda=1$ (critical regime), if the system size is considered as a Fibonacci number (we have considered here $b$ as the ratio of two Fibonacci numbers), a sub-diffusive scaling $N^{-1.26}$ is observed  whereas for the case of non-Fibonacci numbers the sub-diffusive scaling $N^{-1.4}$ is observed. It is interesting to see that, in presence of probes, for both cases, a crossover from sub-diffusive to diffusive regime is observed, as shown in Fig.~\ref{figx}. Furthermore we observe that for all $\lambda$, interestingly, the crossover length from coherent to diffusive transport is independent of the temperature, as displayed in Fig.~\ref{fig5a}.

Since the transport eventually becomes diffusive in presence of probes, in the diffusive regime we define the electrical conductivity $\sigma = \lim_{N \to \infty} \sigma(N)$ with $\sigma(N)= N G$ being the finite-size conductivity. 
In Fig.~\ref{fig6}, we display $\sigma$ as a function of $\lambda$, the strength of the quasi-periodic potential.
 The conductivity $\sigma$ strongly depends on $\lambda$ with the value monotonically decreasing with increasing $\lambda$. This implies that even though the system is in the diffusive regime, the conductivity is still larger in $\lambda<1$ regime in comparison to $\lambda>1$ regime.
However, for large $\gamma$, the quasi-periodic nature of the lattice is smeared out by the probe coupling and conductivity becomes essentially insensitive to the value of $\lambda$.



\begin{figure}[h]
\includegraphics[width=1.0\columnwidth]{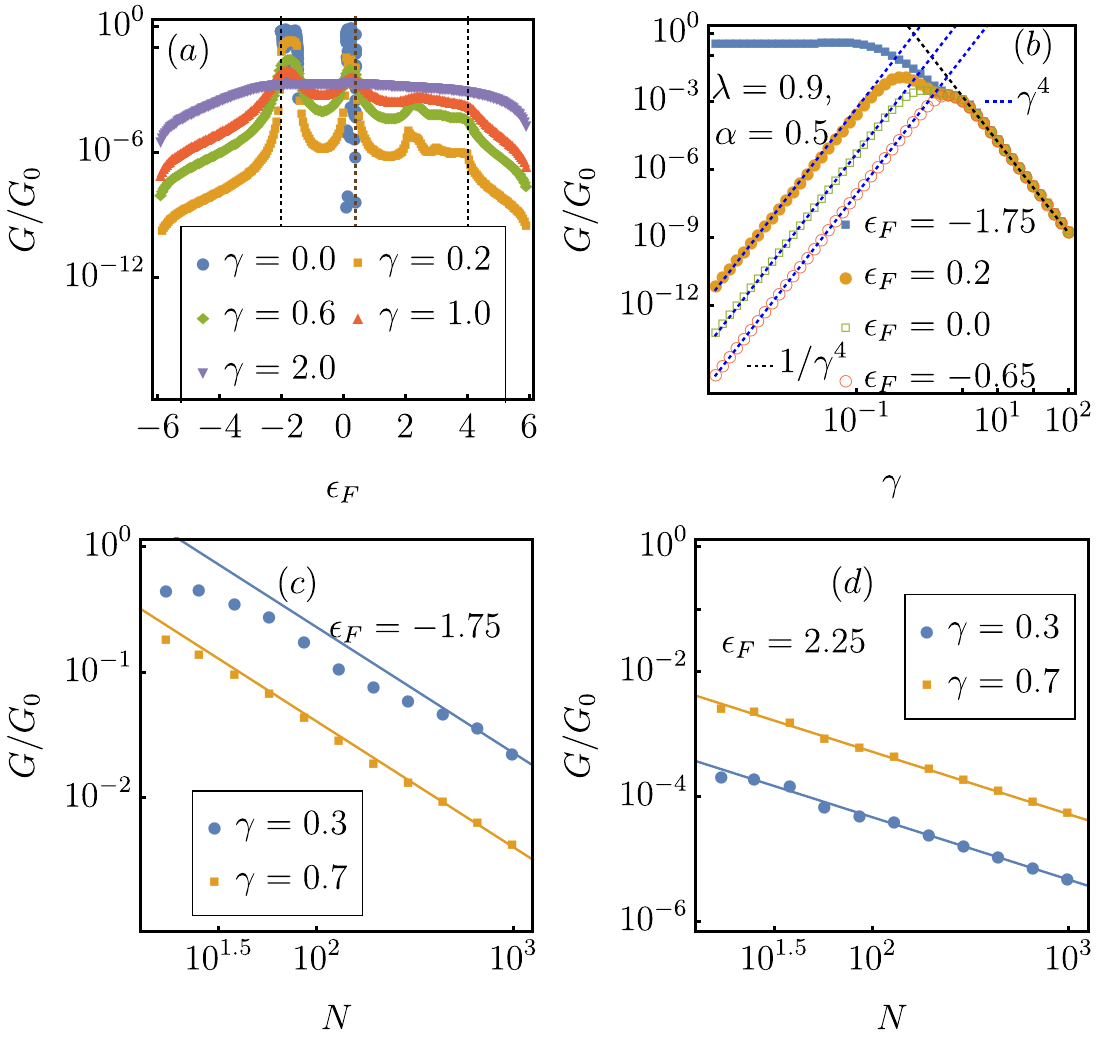} 
\caption{(Color online): Results for the GAAH model: Plot (a) displays the effect of probes on transport. The dotted vertical lines at the two ends represent the band edges and the vertical line in the middle represents the mobility edge. Plot (b) displays conductance as a function of $\gamma$ for different Fermi energies.  
Plots (c)-(d) show the crossover to diffusive regime from both no-transport and transport regimes in presence of probes.}
\label{fig7} 
\end{figure}

We next briefly discuss the results for the generalized AAH i.e., the GAAH model (see Eq.~(\ref{onsite})).  In this case, we observe similar trends in conductance to the one observed for the AAH model. As done in Fig.~\ref{fig1}, in Fig.~\ref{fig7}, we plot $G$ as a function of $\epsilon_F$ in presence of mobility edge. 
For the chosen parameters in our simulations, all the states above (below) the mobility edge are localized (delocalized).  As can be seen in Fig.~\ref{fig7}(b), in all the no-transport regimes, due to the coupling with probes, there is enhancement of $G$  up to a critical $\gamma_c \sim t, \gamma_L, \gamma_R$ with conductance increasing as $\gamma^4$ and thus displaying universality. Furthermore, beyond $\gamma_c$, similar to the AAH case, the $G$ decreases as $1/\gamma^4$. In Fig.~\ref{fig7}(c) and (d), we show that for this model also, the transport eventually becomes diffusive at all regimes.

\section{Results- Voltage-temperature probe technique and thermoelectric performance}
In this section we extend our study to voltage-temperature probe that enables us to investigate thermoelectric properties. In this case, the left and right  reservoirs are kept at fixed chemical potentials $(\mu_L, \mu_R)$ and temperatures  $(T_L, T_R)$. Next we impose the voltage-temperature probe conditions of zero particle and heat currents flowing out of each probe and thereby determine the chemical potential and temperature of each probe. Thus, we set,
\begin{align}
I_n&=\frac{e}{\hbar} \sum_{\alpha} \int_{-\infty}^{\infty} d\epsilon \mathcal{T}_{n\alpha}(\epsilon) (f_n (\epsilon)-f_{\alpha} (\epsilon))=0,  \nonumber \\
Q_n&= \frac{1}{\hbar} \sum_{\alpha} \int_{-\infty}^{\infty} d\epsilon (\epsilon-\mu_n) \mathcal{T}_{n\alpha}(\epsilon) (f_n (\epsilon)-f_{\alpha} (\epsilon))=0. \nonumber \\
&\quad \quad n= 1, 2, \cdots, N
\label{condition1} 
\end{align}
Once again, we focus on the linear response regime and obtain $\mu_n$ and $T_n$ for the probes in terms of $\mu_L, \mu_R$ $T_L$, and $T_R$. Finally, we can write the particle ($I_R$) and heat currents ($Q_R$) flowing out from the right bath as,
\begin{align}
\la I_R \ra&= G\,  \Delta V + G\, S \,\Delta T  \nonumber\\ 
\la Q_R \ra&= G\, \Pi \Delta V + (K + G \,S \, \Pi) \Delta T
\label{lrthermo}
\end{align}
\begin{figure*}
\includegraphics[width=2.0\columnwidth]{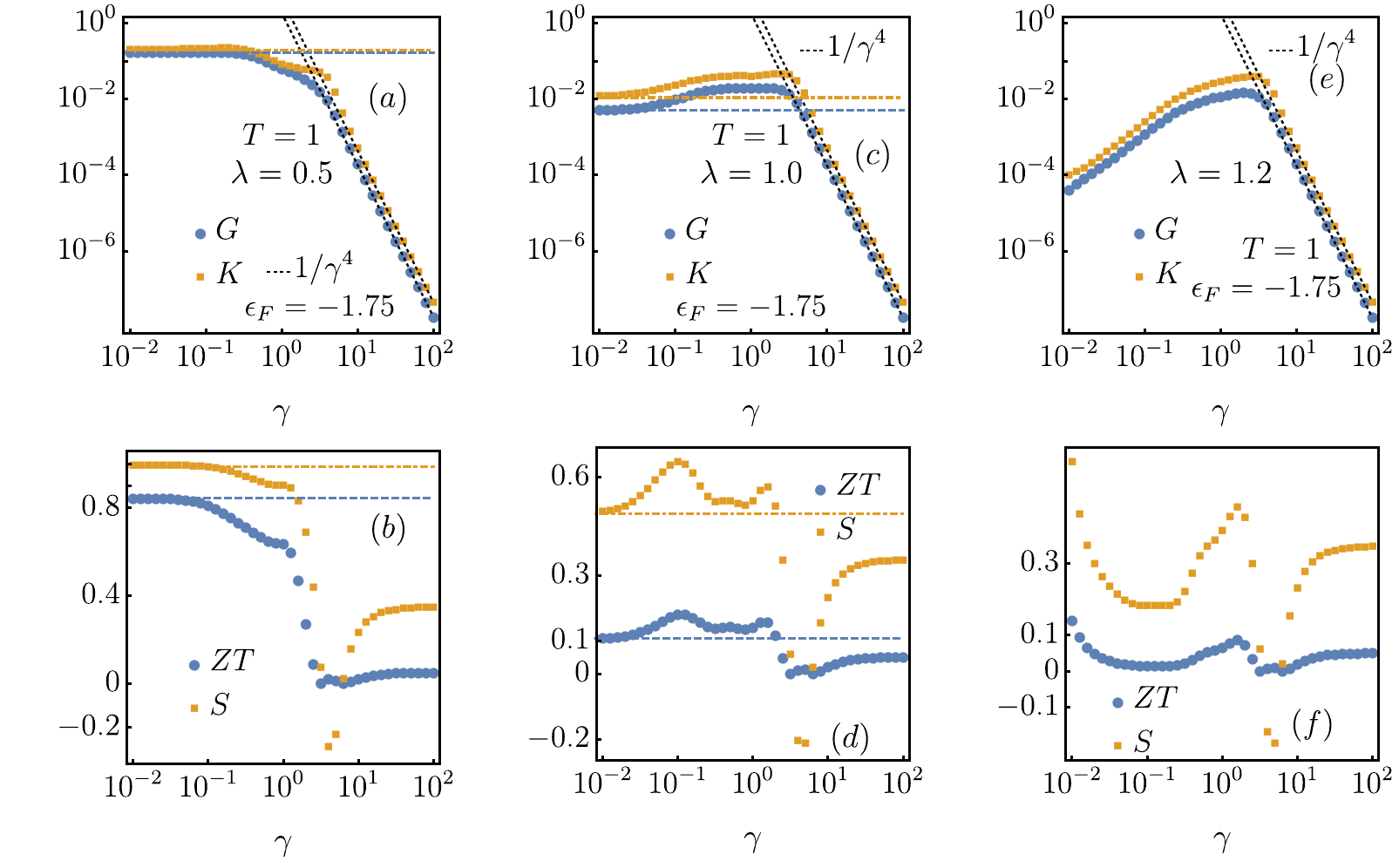} 
\caption{(Color online): Plots (a)-(f) display the various transport coefficients and the ZT for the AAH model as a function of probe coupling strength $\gamma$ in presence of voltage-temperature probes. The horizontal lines in (a)-(d) represent the values of transport coefficients in the coherent limit ($\gamma=0$). }
\label{fig11a}  
\end{figure*}
\begin{figure}[h]
\includegraphics[width=\columnwidth]{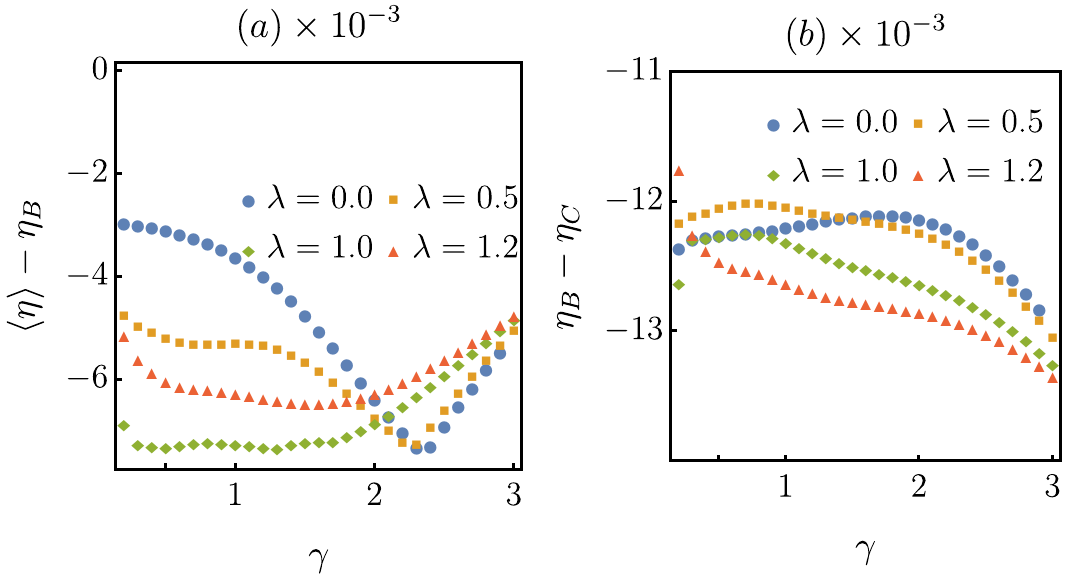} 
\caption{(Color online):  Plots for (a) tighter bound on mean efficiency i.e.,$ \la \eta \ra - \eta_B$ and (b) the bound $\eta_B - \eta_C$ as functions of $\gamma$ for different values of $\lambda$. The parameters are $N=34$, $T=1, \Delta T= 0.2$,  $\mu=-1.75$, $\mu_R- \mu_L= -0.001$.}
\label{fig12} 
\end{figure}
where $G$ is the electrical conductance, $K$ is the thermal conductance, $S$ is the Seebeck coefficient and $\Pi$ is the Peltier coefficient. Here $\Delta V= (\mu_R\!-\!\mu_L)/e$ and $\Delta T = T_R - T_L$.  All of the above transport coefficients depend on the average temperature $T= (T_L+ T_R)/2$ and average chemical potential $\epsilon_F=(\mu_L+\mu_R)/2$. In presence of the probes, we numerically verified the Onsager's reciprocity relation, given by $\Pi = T S$. Note that in order to compute $G$, $\Pi$, $S$ and $K$ in presence of voltage-temperature probes, we first consider $\Delta T=0$ and extract the transport coefficients $G$ and $\Pi$ and next consider $\Delta V=0$ to extract the other two coefficients $S$ and $K$.  
Now in order to realize a thermoelectric engine, we set $\mu_L > \mu_R$ and $T_R> T_L$ and demand $\la I_R\ra>0$ and $\la Q_R\ra>0$ i.e., heat absorbed from the hot right bath induces particle current against  the chemical potential difference. The average power output is then given as  
 $\langle P \rangle =- \Delta V\, \la I_R \ra$ and as a result, the average thermoelectric efficiency of the engine is given by $\langle \eta \rangle=\frac{\langle P\rangle }{\langle Q_R\rangle}$.  In the linear response regime, the maximum thermoelectric efficiency $\la \eta \ra_{\rm max}$ is characterised by a single dimensionless quantity, the thermoelectric figure of merit $ZT=\frac{G S^2}{K} T$ and is given as,
 \begin{equation}
 \la \eta \ra_{\rm max} = \eta_C \frac{ \sqrt{ZT+1} - 1}{\sqrt{ZT+1} + 1}
 \end{equation}
where $\eta_C= \Delta T/T$ is the Carnot efficiency in the linear response regime. In Fig.~\ref{fig11a} (a)-(f), we display the behaviour of various transport coefficients and $ZT$ for the AAH model as  a function of $\gamma$, for different values of $\lambda$. Like the electrical conductance $G$, the thermal conductance $K$ also shows a similar scaling $1/\gamma^4$ at strong probe coupling in all regimes. The Seebeck coefficient however does not display any such scaling. Starting in the $\lambda<1$ regime, the values for both $G$ and $K$ are always lower than that in the coherent limit, whereas probe induced enhancement can be seen in both $\lambda=1$ and $\lambda>1$ regimes (Fig.~\ref{fig11a} (c)-(e)). As a result, $ZT$ decays with increasing $\gamma$ in the $\lambda<1$ regime but in the other two regimes an enhancement in $ZT$ value, compared to the coherent limit, is seen for some values of $\gamma$. 
Note that the overall value of $ZT$ however always remains higher for $\lambda<1$ in comparison to the other two regimes. For large $\gamma$, all three regimes converge to the same $ZT$ value. 



We next move on to analyze two recently obtained universal bounds \cite{Udo-TUR1,Saryal-bound} for the mean efficiency in the context of our thermoelectric engines. Interestingly, these universal bounds are found to be tighter than the seminal Carnot bound and are written in terms of the current fluctuations. One of the bounds emerge from the recently discovered thermodynamic uncertainty relations (TURs) \cite{Udo-TUR1,TUR-udo-first,TUR-Gingrich} which is a trade-off relation between power, efficiency and power fluctuation. In the context of thermoelectric engine, operating in the linear response regime, following the TURs  we obtain the bound on mean efficiency as, 
\begin{eqnarray}
\la \eta \ra &\leq &  \frac{\mu_L- \mu_R}{eT} \frac{1}{S} \leq \eta_c.
\label{TUR}
\end{eqnarray}
Note that, the Seebeck coefficient $S$ is always positive in the engine regime.

Very recently, another tighter bound on engine's efficiency was obtained in Ref. \cite{Saryal-bound, mohanta} by idenfying the input and output currents and  imposing conditions on the direction of these current such that the thermoelectric setup operates as an engine. Following this bound, we receive,  
\begin{eqnarray}
\la \eta \ra &\leq &  \frac{\mu_L-\mu_R}{eT} \frac{1}{S} \sqrt{\frac{ZT}{ZT+1}} \leq \eta_c.
\label{i-o}
\end{eqnarray}
Since $ZT>0$, we arrive to an important conclusion that the bound predicted by Eq.~(\ref{i-o}) is always tighter than the one given in Eq.~(\ref{TUR}). In Fig.~\ref{fig12}  we asses the bounds in Eq.~(\ref{i-o}) for the AAH model in presence of the probes as a function of $\gamma$ and for different $\lambda$ values.
Denoting $\eta_B =  \frac{\mu_L-\mu_R}{eT} \frac{1}{S} \sqrt{\frac{ZT}{ZT+1}} $, we observe that $\langle \eta \rangle - \eta_B$ and  $\eta_B - \eta_C$ are always negative in the parameter regimes considered here. Thus, we find that the bounds in Eq.~(\ref{TUR}) and Eq.~(\ref{i-o}), obtained in the linear response regime, are valid even in presence of the probes. 


\section{Summary}

In summary, we have investigated the transport properties in quasi-periodic systems following the B{\"u}ttiker probe approach by implementing both   voltage and voltage-temperature probe conditions. Within the voltage-probe framework, in the strong probe coupling limit, we observed a power-law decay of electrical conductance with the probe coupling strength. This behaviour is in fact independent of the Fermi energy, nature of the on-site potential and the  temperature. Moreover, within the voltage-temperature probe scheme, in addition to the electrical conductance, the thermal conductance also display the same power-law decay with the probe coupling. In this sense, the power-law decay of transport coefficients is universal. Importantly, this power law scaling observed here turns out to be different from what was reported in a recent work following the local Lindblad master equation approach \cite{Landi-latest}. This further pin points that, the two different approaches namely NEGF and local Lindblad master equation, effectively mimic different underlying scattering processes. In presence of voltage probe we also observed that the conductance of all the exponentially decaying regimes (namely localized regimes of the AAH model, above the mobility edge of the GAAH model, outside the band edges of the central lattice system, and the band-gap regimes), initially gets enhanced with a particular power-law scaling with the probe coupling strength at zero-temperature. Thus, we have reported here the environment-assisted transport with a particular scaling for the `no transport' regimes at zero temperature. With finite temperature and voltage-temperature probes, no particular scaling is observed in the `no transport' regimes. For sufficiently strong coupling, the transport eventually becomes diffusive in all regimes of the original model. However,  with finite but small probe coupling, a crossover from coherent transport to diffusive transport can be observed. Such crossover length is insensitive to the temperature but depends crucially on the transport regime of the original model.
We also investigated the thermoelectric properties by incorporating voltage-temperature probes and further assed, in presence of probes, the validity of two recently obtained tighter bounds on efficiency and also showed that the bound predicted from the TUR is always looser than the other bound.

\section*{Acknowledgement} The authors would like to acknowledge Archak Purkayastha for numerous useful discussions. M.~S acknowledges funding from National Postdoctoral Fellowship Scheme (NPDF), SERB file No. PDF/2020/000992. B.~P.~V. is supported by  a Department of Science \& Technology Science and Engineering Research Board (India) Start-up Research Grant No.  SRG/2019/001585. BKA acknowledges the MATRICS grant MTR/2020/000472 from SERB, Government of India and the Shastri Indo-Canadian Institute for providing financial support for this research work in the form of a Shastri Institutional Collaborative
Research Grant (SICRG).

\section*{Appendix A: Proof of the universal power-law decay $\frac{1}{N \gamma^4}$ for the conductance $G$ in the strong $\gamma$ limit}
In this section we provide a rigorous proof for the universal power-law decay of the conductance with the system size and the probe coupling strength $\gamma$  in the strong probe coupling limit and at zero temperature. We show that in this limit, the conductance takes the form $G \sim \frac{1}{N \gamma^4}$ which further indicates a diffusive behaviour, as expected in the strong probe coupling limit.  Now in order to arrive at this result, we need to compute the effective transmission, as given in Eq.~(\ref{eff-trans}). We therefore first compute the retarded Green's function components in strong $\gamma$ limit followed by the calculation of the ${\cal W}$ matrix. 
\subsubsection{Analytical calculation of the retarded Green's function ${\cal G}^r_{ij}$ in the strong $\gamma$ limit}
In general, for any nearest neighbour tight-binding lattice with onsite energy $\epsilon_i$ and hopping strength $t$, we can write down the the retarded Green's function ${\cal G}^r$ as,
${\cal G}^r=\mathcal{M}^{-1}=\frac{1}{t} \begin{pmatrix}
a_{11} & 1 & 0 & 0 & \ldots \\
1& a_{22} & 1 & 0 & \ldots \\
0 & 1 & a_{33} &1 & \ldots \\
\ldots & \ldots & \ldots & \ldots \\
0 & 0 & 0 & 1 & a_{NN}
\end{pmatrix}^{-1}$. 
Note that the first and the last site of the lattice are coupled to the left and the right reservoir, respectively, and also to the probes, whereas, all other sites are only connected to the probes.  As a result,  we can write down the matrix elements as
\begin{eqnarray}
a_{11} &=&\frac{1}{t} \Big[\epsilon \!-\!\epsilon_{1} \!-\!\frac{(\gamma_{L}^2+\gamma_P^2)}{2 t_0^2} \Big[\ep \!-\!i \sqrt{4 t_0^2 \!-\! \ep^2}\Big]\Big], \nonumber \\
a_{2i} &=&\frac{1}{t} \Big[\epsilon - \epsilon_i-\frac{\gamma_P^2}{2 t_0^2} \Big[\ep \!-\!i \sqrt{\!4 t_0^2-\ep^2\!}\Big]\Big], i=2,\cdots N-1, \nonumber \\
a_{NN}&=&\frac{1}{t} \Big[\epsilon -\epsilon_{N} -\frac{(\gamma_{R}^2+\gamma_P^2)}{2 t_0^2} \Big[\ep \!-\!i \sqrt{\!4 t_0^2-\ep^2\!}\Big]\Big]. 
\end{eqnarray}
where we used the self-energy expression $\Sigma^r_{\alpha}(\ep)=\frac{\gamma_{\alpha}^2}{2\, t^{2}_0}\Big[\ep \!-\!i \sqrt{4 t_0^2\!-\! \ep^2}\Big], \quad \alpha=L, R, P.$  Now, as done in our numerics, we assume homogenous probe coupling strength i.e., $\gamma_P=\gamma$. In the limit of strong coupling $\gamma \gg t_0, \epsilon_i, \epsilon, \gamma_L, \gamma_R$, all the diagonal elements of the above matrix are the same and given as $a_{ii}=-i \frac{\gamma^2}{t \, t_0}, i=1, 2, \cdots N$. Note that to arrive at this result we also assumed $t_0$ to be larger than system band energies. In what follows, since we are only interested in the scaling with $\gamma^2$, we therefore only keep track of this parameters and suppress all the other parameters of the setup. 
 
The inverse of this tridiagonal matrix can be computed easily following a transfer matrix approach \cite{Roy-1}. It is to check that 
\begin{equation}
{\cal G}^r_{ij}=(-1)^{i+j} \frac{\Delta_{1, i-1} \Delta_{N-j, N}}{\Delta_{1, N}}, \quad j > i,
\end{equation} 
and ${\cal G}^r_{ij}={\cal G}^r_{ji}$ for $i >j$. Here $\Delta_{i,j}$ is the determinant of the sub-matrix starting with the $i-$th row and column and ending with $j-$th row and column. The determinant of each of this sub-matrix is related to the product of transfer matrices 
$T_a=\begin{pmatrix}
-i \gamma^2 & -1 \\
1 & 0
\end{pmatrix}$.
With this in hand, it is easy to check that ${\cal G}^r_{ij} \sim |\lambda_{+}|^{-|i-j|-1}$ where $\lambda_{+}$ is the largest eigenvalue of the $2\times 2$ transfer matrix $T_a$. In this strong $\gamma$ limit, $|\lambda_{+}|\sim \gamma^2$. Thus, ${\cal G}^r_{ij}\sim \gamma^{-2(|j-i|+1)}$. Similarly, ${\cal G}_{iN} \sim \gamma^{-2(|N-i|+1)}$ and ${\cal G}^r_{1i}\sim\gamma^{-2 i}$. As a result,  we can write down the transmission probabilities as, $T_{ij}=\gamma^4 |{\cal G}_{ij}|^2\sim \gamma^{-4|j-i|} $, $T_{iR}=T_{Ri}=\gamma^2 |{\cal G}^r_{iN}|^2 \sim \gamma^{-2} \gamma^{-4|N-i|}$ and $T_{iL}=T_{Li}=\gamma^2 |{\cal G}_{i1}|^2 \sim \gamma^{-2} \gamma^{-4|i-1|}$.

\subsubsection{Analytical form of the $\mathcal{W}$ matrix and its inverse in the strong $\gamma$ limit}
With the above result for ${\cal G}_{ij}$, we now construct the ${\cal W}$ matrix in the strong $\gamma$ regime.  Along the diagonal of the $\mathcal{W}$ matrix, the first and last matrix elements contain the dominant contribution as $1/\gamma^2$ which appears from $T_{1L}$ and $T_{NR}$ terms, respectively. The other diagonal elements depend on $\gamma$ as $1/\gamma^4$.   The off-diagonal elements of ${\cal W}_{ij}$ depend on $\gamma$ as $\gamma^{-4 |i-j|}$. We therefore consider only the first neighbours of the diagonal entries i.e., for  $|i-j|=1$ which gives the dominant contribution as $1/\gamma^4$ and all the other off-diagonal terms can be neglected. Thus, $\mathcal{W}$  reduces to a tridiagonal matrix in this strong $\gamma$ limit, given as, 
$\mathcal{W}=\begin{pmatrix}
1/\gamma^2 & -1/\gamma^4 & 0 & 0 & \ldots \\
-1/\gamma^4 & 2/\gamma^4 & -1/\gamma^4 & 0 & \ldots \\
0 & -1/\gamma^4 & 2/\gamma^4 & -1/\gamma^4 & \ldots \\
\ldots & \ldots & \ldots & \ldots \\
0 & 0 & 0 & -1/\gamma^4 & 1/\gamma^2.
\end{pmatrix}$
Now to check the leading order behaviour of $\gamma$ in the conductance following the expression for effective transmission (Eq.~(\ref{eff-trans})) the most dominant contribution comes from $T_{NR} \sim 1/\gamma^2$ (independent of $N$) and $T_{1L}\sim 1/\gamma^2$ (independent of $N$). Thus, we need to calculate $\mathcal{W}^{-1}_{N1}$ element to see the leading order behaviour, given as $\mathcal{W}^{-1}_{N1}=1/ \Delta_{1, N}[{\cal W}]$. The determinant $\Delta_{1,N}[\mathcal{W}]$ can be once again computed following the transfer matrix approach,
\begin{align}
\begin{pmatrix}
\Delta_{1,N} [\mathcal{W}] \\
\Delta_{1,N-1} [\mathcal{W}]
\end{pmatrix}= \frac{1}{\gamma^4} \begin{pmatrix}
\gamma^2 & -1 \\
1 & 0
\end{pmatrix} \begin{pmatrix}
2 & -1 \\
1 & 0
\end{pmatrix}^{N-2} \begin{pmatrix}
\gamma^2 & -1 \\
1 & 0
\end{pmatrix} \begin{pmatrix}
 1 \\
 0
\end{pmatrix}
\end{align}
Interestingly, the matrix $\begin{pmatrix}
2 & -1 \\
1 & 0
\end{pmatrix}$ is non-diagonalizable. One can  however write it as Jordon normal form $J=\begin{pmatrix}
1 & 1 \\
0 & 1
\end{pmatrix}$ using a transformation $S$. Thus, we can rewrite the matrix equation as,
\begin{align}
\begin{pmatrix}
\Delta_{1,N} [\mathcal{W}] \\
\Delta_{1,N-1} [\mathcal{W}]
\end{pmatrix}&= \frac{1}{\gamma^4}\begin{pmatrix}
\gamma^2 & -1 \\
1 & 0
\end{pmatrix} S J^{N-2} S^{-1} \begin{pmatrix}
 \gamma^2 \\
 1
\end{pmatrix} \\ \nonumber
&= \frac{1}{\gamma^4} \begin{pmatrix}
\gamma^2 & -1 \\
1 & 0
\end{pmatrix} S \begin{pmatrix}
1 & N-2 \\
0 & 1
\end{pmatrix} S^{-1} \begin{pmatrix}
 \gamma^2 \\
 1
\end{pmatrix},
\end{align}
with $S=\begin{pmatrix}
i & 0 \\
i & -i
\end{pmatrix}$. Using this, $\Delta_{1,N} [\mathcal{W}] \sim \frac{N}{\gamma^4} (A \gamma^4 + B \gamma^2 + C)+ D$ with constant $A, B, C$ and $D$. 
Thus with large $\gamma$, $\mathcal{W}^{-1}_{N1}\sim 1/(A N + D)$ and is independent of $\gamma$. Thus, in the second term of Eq.~(\ref{eff-trans}), we obtain,  $ G  \propto T_{\rm eff}\approx T_{RN}\mathcal{W}^{-1}_{N1} T_{1L} \sim \frac{1}{N \gamma^4}$ which gives us the desired scaling for conductance. The derived result also shows that in the strong $\gamma$ regime the conductance scales as $1/N$, indicating a diffusive behaviour.


\bibliography{ref_probe}

\begin{thebibliography}{65}%
\makeatletter
\providecommand \@ifxundefined [1]{%
 \@ifx{#1\undefined}
}%
\providecommand \@ifnum [1]{%
 \ifnum #1\expandafter \@firstoftwo
 \else \expandafter \@secondoftwo
 \fi
}%
\providecommand \@ifx [1]{%
 \ifx #1\expandafter \@firstoftwo
 \else \expandafter \@secondoftwo
 \fi
}%
\providecommand \natexlab [1]{#1}%
\providecommand \enquote  [1]{``#1''}%
\providecommand \bibnamefont  [1]{#1}%
\providecommand \bibfnamefont [1]{#1}%
\providecommand \citenamefont [1]{#1}%
\providecommand \href@noop [0]{\@secondoftwo}%
\providecommand \href [0]{\begingroup \@sanitize@url \@href}%
\providecommand \@href[1]{\@@startlink{#1}\@@href}%
\providecommand \@@href[1]{\endgroup#1\@@endlink}%
\providecommand \@sanitize@url [0]{\catcode `\\12\catcode `\$12\catcode
  `\&12\catcode `\#12\catcode `\^12\catcode `\_12\catcode `\%12\relax}%
\providecommand \@@startlink[1]{}%
\providecommand \@@endlink[0]{}%
\providecommand \url  [0]{\begingroup\@sanitize@url \@url }%
\providecommand \@url [1]{\endgroup\@href {#1}{\urlprefix }}%
\providecommand \urlprefix  [0]{URL }%
\providecommand \Eprint [0]{\href }%
\providecommand \doibase [0]{http://dx.doi.org/}%
\providecommand \selectlanguage [0]{\@gobble}%
\providecommand \bibinfo  [0]{\@secondoftwo}%
\providecommand \bibfield  [0]{\@secondoftwo}%
\providecommand \translation [1]{[#1]}%
\providecommand \BibitemOpen [0]{}%
\providecommand \bibitemStop [0]{}%
\providecommand \bibitemNoStop [0]{.\EOS\space}%
\providecommand \EOS [0]{\spacefactor3000\relax}%
\providecommand \BibitemShut  [1]{\csname bibitem#1\endcsname}%
\let\auto@bib@innerbib\@empty
\bibitem [{\citenamefont {Aubry}\ and\ \citenamefont
  {Andr{\'e}}(1980)}]{Aubry2}%
  \BibitemOpen
  \bibfield  {author} {\bibinfo {author} {\bibfnamefont {S.}~\bibnamefont
  {Aubry}}\ and\ \bibinfo {author} {\bibfnamefont {G.}~\bibnamefont
  {Andr{\'e}}},\ }\href@noop {} {\bibfield  {journal} {\bibinfo  {journal}
  {Ann. Israel Phys. Soc.}\ }\textbf {\bibinfo {volume} {3}},\ \bibinfo {pages}
  {133} (\bibinfo {year} {1980})}\BibitemShut {NoStop}%
\bibitem [{\citenamefont {Harper}(1955)}]{Harper}%
  \BibitemOpen
  \bibfield  {author} {\bibinfo {author} {\bibfnamefont {P.~G.}\ \bibnamefont
  {Harper}},\ }\href {\doibase 10.1088/0370-1298/68/10/304} {\bibfield
  {journal} {\bibinfo  {journal} {Proceedings of the Physical Society. Section
  A}\ }\textbf {\bibinfo {volume} {68}},\ \bibinfo {pages} {874} (\bibinfo
  {year} {1955})}\BibitemShut {NoStop}%
\bibitem [{\citenamefont {Ganeshan}\ \emph {et~al.}(2013)\citenamefont
  {Ganeshan}, \citenamefont {Sun},\ and\ \citenamefont
  {Das~Sarma}}]{Commensurate_AAH}%
  \BibitemOpen
  \bibfield  {author} {\bibinfo {author} {\bibfnamefont {S.}~\bibnamefont
  {Ganeshan}}, \bibinfo {author} {\bibfnamefont {K.}~\bibnamefont {Sun}}, \
  and\ \bibinfo {author} {\bibfnamefont {S.}~\bibnamefont {Das~Sarma}},\ }\href
  {\doibase 10.1103/PhysRevLett.110.180403} {\bibfield  {journal} {\bibinfo
  {journal} {Phys. Rev. Lett.}\ }\textbf {\bibinfo {volume} {110}},\ \bibinfo
  {pages} {180403} (\bibinfo {year} {2013})}\BibitemShut {NoStop}%
\bibitem [{\citenamefont {Ganeshan}\ \emph {et~al.}(2015)\citenamefont
  {Ganeshan}, \citenamefont {Pixley},\ and\ \citenamefont
  {Das~Sarma}}]{GAAH_mobility_edge}%
  \BibitemOpen
  \bibfield  {author} {\bibinfo {author} {\bibfnamefont {S.}~\bibnamefont
  {Ganeshan}}, \bibinfo {author} {\bibfnamefont {J.~H.}\ \bibnamefont
  {Pixley}}, \ and\ \bibinfo {author} {\bibfnamefont {S.}~\bibnamefont
  {Das~Sarma}},\ }\href {\doibase 10.1103/PhysRevLett.114.146601} {\bibfield
  {journal} {\bibinfo  {journal} {Phys. Rev. Lett.}\ }\textbf {\bibinfo
  {volume} {114}},\ \bibinfo {pages} {146601} (\bibinfo {year}
  {2015})}\BibitemShut {NoStop}%
\bibitem [{\citenamefont {Ostlund}\ \emph
  {et~al.}(1983{\natexlab{a}})\citenamefont {Ostlund}, \citenamefont {Pandit},
  \citenamefont {Rand}, \citenamefont {Schellnhuber},\ and\ \citenamefont
  {Siggia}}]{Pandit83}%
  \BibitemOpen
  \bibfield  {author} {\bibinfo {author} {\bibfnamefont {S.}~\bibnamefont
  {Ostlund}}, \bibinfo {author} {\bibfnamefont {R.}~\bibnamefont {Pandit}},
  \bibinfo {author} {\bibfnamefont {D.}~\bibnamefont {Rand}}, \bibinfo {author}
  {\bibfnamefont {H.~J.}\ \bibnamefont {Schellnhuber}}, \ and\ \bibinfo
  {author} {\bibfnamefont {E.~D.}\ \bibnamefont {Siggia}},\ }\href {\doibase
  10.1103/PhysRevLett.50.1873} {\bibfield  {journal} {\bibinfo  {journal}
  {Phys. Rev. Lett.}\ }\textbf {\bibinfo {volume} {50}},\ \bibinfo {pages}
  {1873} (\bibinfo {year} {1983}{\natexlab{a}})}\BibitemShut {NoStop}%
\bibitem [{\citenamefont {Sutradhar}\ \emph {et~al.}(2019)\citenamefont
  {Sutradhar}, \citenamefont {Mukerjee}, \citenamefont {Pandit},\ and\
  \citenamefont {Banerjee}}]{Sumilan}%
  \BibitemOpen
  \bibfield  {author} {\bibinfo {author} {\bibfnamefont {J.}~\bibnamefont
  {Sutradhar}}, \bibinfo {author} {\bibfnamefont {S.}~\bibnamefont {Mukerjee}},
  \bibinfo {author} {\bibfnamefont {R.}~\bibnamefont {Pandit}}, \ and\ \bibinfo
  {author} {\bibfnamefont {S.}~\bibnamefont {Banerjee}},\ }\href {\doibase
  10.1103/PhysRevB.99.224204} {\bibfield  {journal} {\bibinfo  {journal} {Phys.
  Rev. B}\ }\textbf {\bibinfo {volume} {99}},\ \bibinfo {pages} {224204}
  (\bibinfo {year} {2019})}\BibitemShut {NoStop}%
\bibitem [{\citenamefont {Modak}\ and\ \citenamefont {Mukerjee}(2015)}]{Modak}%
  \BibitemOpen
  \bibfield  {author} {\bibinfo {author} {\bibfnamefont {R.}~\bibnamefont
  {Modak}}\ and\ \bibinfo {author} {\bibfnamefont {S.}~\bibnamefont
  {Mukerjee}},\ }\href {\doibase 10.1103/PhysRevLett.115.230401} {\bibfield
  {journal} {\bibinfo  {journal} {Phys. Rev. Lett.}\ }\textbf {\bibinfo
  {volume} {115}},\ \bibinfo {pages} {230401} (\bibinfo {year}
  {2015})}\BibitemShut {NoStop}%
\bibitem [{\citenamefont {Choi}\ \emph {et~al.}(2016)\citenamefont {Choi},
  \citenamefont {Hild}, \citenamefont {Zeiher}, \citenamefont {Schau{\ss}},
  \citenamefont {Rubio-Abadal}, \citenamefont {Yefsah}, \citenamefont
  {Khemani}, \citenamefont {Huse}, \citenamefont {Bloch},\ and\ \citenamefont
  {Gross}}]{I_bloch_experiment}%
  \BibitemOpen
  \bibfield  {author} {\bibinfo {author} {\bibfnamefont {J.-y.}\ \bibnamefont
  {Choi}}, \bibinfo {author} {\bibfnamefont {S.}~\bibnamefont {Hild}}, \bibinfo
  {author} {\bibfnamefont {J.}~\bibnamefont {Zeiher}}, \bibinfo {author}
  {\bibfnamefont {P.}~\bibnamefont {Schau{\ss}}}, \bibinfo {author}
  {\bibfnamefont {A.}~\bibnamefont {Rubio-Abadal}}, \bibinfo {author}
  {\bibfnamefont {T.}~\bibnamefont {Yefsah}}, \bibinfo {author} {\bibfnamefont
  {V.}~\bibnamefont {Khemani}}, \bibinfo {author} {\bibfnamefont {D.~A.}\
  \bibnamefont {Huse}}, \bibinfo {author} {\bibfnamefont {I.}~\bibnamefont
  {Bloch}}, \ and\ \bibinfo {author} {\bibfnamefont {C.}~\bibnamefont
  {Gross}},\ }\href {\doibase 10.1126/science.aaf8834} {\bibfield  {journal}
  {\bibinfo  {journal} {Science}\ }\textbf {\bibinfo {volume} {352}},\ \bibinfo
  {pages} {1547} (\bibinfo {year} {2016})}\BibitemShut {NoStop}%
\bibitem [{\citenamefont {Kraus}\ \emph {et~al.}(2012)\citenamefont {Kraus},
  \citenamefont {Lahini}, \citenamefont {Ringel}, \citenamefont {Verbin},\ and\
  \citenamefont {Zilberberg}}]{zilberberg1}%
  \BibitemOpen
  \bibfield  {author} {\bibinfo {author} {\bibfnamefont {Y.~E.}\ \bibnamefont
  {Kraus}}, \bibinfo {author} {\bibfnamefont {Y.}~\bibnamefont {Lahini}},
  \bibinfo {author} {\bibfnamefont {Z.}~\bibnamefont {Ringel}}, \bibinfo
  {author} {\bibfnamefont {M.}~\bibnamefont {Verbin}}, \ and\ \bibinfo {author}
  {\bibfnamefont {O.}~\bibnamefont {Zilberberg}},\ }\href {\doibase
  10.1103/PhysRevLett.109.106402} {\bibfield  {journal} {\bibinfo  {journal}
  {Phys. Rev. Lett.}\ }\textbf {\bibinfo {volume} {109}},\ \bibinfo {pages}
  {106402} (\bibinfo {year} {2012})}\BibitemShut {NoStop}%
\bibitem [{\citenamefont {Verbin}\ \emph {et~al.}(2013)\citenamefont {Verbin},
  \citenamefont {Zilberberg}, \citenamefont {Kraus}, \citenamefont {Lahini},\
  and\ \citenamefont {Silberberg}}]{zilberberg2}%
  \BibitemOpen
  \bibfield  {author} {\bibinfo {author} {\bibfnamefont {M.}~\bibnamefont
  {Verbin}}, \bibinfo {author} {\bibfnamefont {O.}~\bibnamefont {Zilberberg}},
  \bibinfo {author} {\bibfnamefont {Y.~E.}\ \bibnamefont {Kraus}}, \bibinfo
  {author} {\bibfnamefont {Y.}~\bibnamefont {Lahini}}, \ and\ \bibinfo {author}
  {\bibfnamefont {Y.}~\bibnamefont {Silberberg}},\ }\href {\doibase
  10.1103/PhysRevLett.110.076403} {\bibfield  {journal} {\bibinfo  {journal}
  {Phys. Rev. Lett.}\ }\textbf {\bibinfo {volume} {110}},\ \bibinfo {pages}
  {076403} (\bibinfo {year} {2013})}\BibitemShut {NoStop}%
\bibitem [{\citenamefont {An}\ \emph {et~al.}(2020)\citenamefont {An},
  \citenamefont {Padavić}, \citenamefont {Meier}, \citenamefont {Hegde},
  \citenamefont {Ganeshan}, \citenamefont {Pixley}, \citenamefont
  {Vishveshwara},\ and\ \citenamefont {Gadway}}]{GAAH_experiment}%
  \BibitemOpen
  \bibfield  {author} {\bibinfo {author} {\bibfnamefont {F.~A.}\ \bibnamefont
  {An}}, \bibinfo {author} {\bibfnamefont {K.}~\bibnamefont {Padavić}},
  \bibinfo {author} {\bibfnamefont {E.~J.}\ \bibnamefont {Meier}}, \bibinfo
  {author} {\bibfnamefont {S.}~\bibnamefont {Hegde}}, \bibinfo {author}
  {\bibfnamefont {S.}~\bibnamefont {Ganeshan}}, \bibinfo {author}
  {\bibfnamefont {J.~H.}\ \bibnamefont {Pixley}}, \bibinfo {author}
  {\bibfnamefont {S.}~\bibnamefont {Vishveshwara}}, \ and\ \bibinfo {author}
  {\bibfnamefont {B.}~\bibnamefont {Gadway}},\ }\href@noop {} {\  (\bibinfo
  {year} {2020})},\ \Eprint {http://arxiv.org/abs/2007.01393}
  {arXiv:2007.01393} \BibitemShut {NoStop}%
\bibitem [{\citenamefont {L\"uschen}\ \emph {et~al.}(2018)\citenamefont
  {L\"uschen}, \citenamefont {Scherg}, \citenamefont {Kohlert}, \citenamefont
  {Schreiber}, \citenamefont {Bordia}, \citenamefont {Li}, \citenamefont
  {Das~Sarma},\ and\ \citenamefont {Bloch}}]{expt1}%
  \BibitemOpen
  \bibfield  {author} {\bibinfo {author} {\bibfnamefont {H.~P.}\ \bibnamefont
  {L\"uschen}}, \bibinfo {author} {\bibfnamefont {S.}~\bibnamefont {Scherg}},
  \bibinfo {author} {\bibfnamefont {T.}~\bibnamefont {Kohlert}}, \bibinfo
  {author} {\bibfnamefont {M.}~\bibnamefont {Schreiber}}, \bibinfo {author}
  {\bibfnamefont {P.}~\bibnamefont {Bordia}}, \bibinfo {author} {\bibfnamefont
  {X.}~\bibnamefont {Li}}, \bibinfo {author} {\bibfnamefont {S.}~\bibnamefont
  {Das~Sarma}}, \ and\ \bibinfo {author} {\bibfnamefont {I.}~\bibnamefont
  {Bloch}},\ }\href {\doibase 10.1103/PhysRevLett.120.160404} {\bibfield
  {journal} {\bibinfo  {journal} {Phys. Rev. Lett.}\ }\textbf {\bibinfo
  {volume} {120}},\ \bibinfo {pages} {160404} (\bibinfo {year}
  {2018})}\BibitemShut {NoStop}%
\bibitem [{\citenamefont {L\"uschen}\ \emph {et~al.}(2017)\citenamefont
  {L\"uschen}, \citenamefont {Bordia}, \citenamefont {Hodgman}, \citenamefont
  {Schreiber}, \citenamefont {Sarkar}, \citenamefont {Daley}, \citenamefont
  {Fischer}, \citenamefont {Altman}, \citenamefont {Bloch},\ and\ \citenamefont
  {Schneider}}]{expt2}%
  \BibitemOpen
  \bibfield  {author} {\bibinfo {author} {\bibfnamefont {H.~P.}\ \bibnamefont
  {L\"uschen}}, \bibinfo {author} {\bibfnamefont {P.}~\bibnamefont {Bordia}},
  \bibinfo {author} {\bibfnamefont {S.~S.}\ \bibnamefont {Hodgman}}, \bibinfo
  {author} {\bibfnamefont {M.}~\bibnamefont {Schreiber}}, \bibinfo {author}
  {\bibfnamefont {S.}~\bibnamefont {Sarkar}}, \bibinfo {author} {\bibfnamefont
  {A.~J.}\ \bibnamefont {Daley}}, \bibinfo {author} {\bibfnamefont {M.~H.}\
  \bibnamefont {Fischer}}, \bibinfo {author} {\bibfnamefont {E.}~\bibnamefont
  {Altman}}, \bibinfo {author} {\bibfnamefont {I.}~\bibnamefont {Bloch}}, \
  and\ \bibinfo {author} {\bibfnamefont {U.}~\bibnamefont {Schneider}},\ }\href
  {\doibase 10.1103/PhysRevX.7.011034} {\bibfield  {journal} {\bibinfo
  {journal} {Phys. Rev. X}\ }\textbf {\bibinfo {volume} {7}},\ \bibinfo {pages}
  {011034} (\bibinfo {year} {2017})}\BibitemShut {NoStop}%
\bibitem [{\citenamefont {Atala}\ \emph {et~al.}(2014)\citenamefont {Atala},
  \citenamefont {Aidelsburger}, \citenamefont {Lohse}, \citenamefont
  {Barreiro}, \citenamefont {Paredes},\ and\ \citenamefont {Bloch}}]{expt3}%
  \BibitemOpen
  \bibfield  {author} {\bibinfo {author} {\bibfnamefont {M.}~\bibnamefont
  {Atala}}, \bibinfo {author} {\bibfnamefont {M.}~\bibnamefont {Aidelsburger}},
  \bibinfo {author} {\bibfnamefont {M.}~\bibnamefont {Lohse}}, \bibinfo
  {author} {\bibfnamefont {J.~T.}\ \bibnamefont {Barreiro}}, \bibinfo {author}
  {\bibfnamefont {B.}~\bibnamefont {Paredes}}, \ and\ \bibinfo {author}
  {\bibfnamefont {I.}~\bibnamefont {Bloch}},\ }\href
  {https://doi.org/10.1038/nphys2998} {\bibfield  {journal} {\bibinfo
  {journal} {Nature Physics}\ }\textbf {\bibinfo {volume} {10}},\ \bibinfo
  {pages} {588} (\bibinfo {year} {2014})}\BibitemShut {NoStop}%
\bibitem [{\citenamefont {Lahini}\ \emph {et~al.}(2009)\citenamefont {Lahini},
  \citenamefont {Pugatch}, \citenamefont {Pozzi}, \citenamefont {Sorel},
  \citenamefont {Morandotti}, \citenamefont {Davidson},\ and\ \citenamefont
  {Silberberg}}]{AAH-expt-Silberberg}%
  \BibitemOpen
  \bibfield  {author} {\bibinfo {author} {\bibfnamefont {Y.}~\bibnamefont
  {Lahini}}, \bibinfo {author} {\bibfnamefont {R.}~\bibnamefont {Pugatch}},
  \bibinfo {author} {\bibfnamefont {F.}~\bibnamefont {Pozzi}}, \bibinfo
  {author} {\bibfnamefont {M.}~\bibnamefont {Sorel}}, \bibinfo {author}
  {\bibfnamefont {R.}~\bibnamefont {Morandotti}}, \bibinfo {author}
  {\bibfnamefont {N.}~\bibnamefont {Davidson}}, \ and\ \bibinfo {author}
  {\bibfnamefont {Y.}~\bibnamefont {Silberberg}},\ }\href {\doibase
  10.1103/PhysRevLett.103.013901} {\bibfield  {journal} {\bibinfo  {journal}
  {Phys. Rev. Lett.}\ }\textbf {\bibinfo {volume} {103}},\ \bibinfo {pages}
  {013901} (\bibinfo {year} {2009})}\BibitemShut {NoStop}%
\bibitem [{\citenamefont {Kohlert}\ \emph {et~al.}(2019)\citenamefont
  {Kohlert}, \citenamefont {Scherg}, \citenamefont {Li}, \citenamefont
  {L\"uschen}, \citenamefont {Das~Sarma}, \citenamefont {Bloch},\ and\
  \citenamefont {Aidelsburger}}]{mobility_edge_expt}%
  \BibitemOpen
  \bibfield  {author} {\bibinfo {author} {\bibfnamefont {T.}~\bibnamefont
  {Kohlert}}, \bibinfo {author} {\bibfnamefont {S.}~\bibnamefont {Scherg}},
  \bibinfo {author} {\bibfnamefont {X.}~\bibnamefont {Li}}, \bibinfo {author}
  {\bibfnamefont {H.~P.}\ \bibnamefont {L\"uschen}}, \bibinfo {author}
  {\bibfnamefont {S.}~\bibnamefont {Das~Sarma}}, \bibinfo {author}
  {\bibfnamefont {I.}~\bibnamefont {Bloch}}, \ and\ \bibinfo {author}
  {\bibfnamefont {M.}~\bibnamefont {Aidelsburger}},\ }\href {\doibase
  10.1103/PhysRevLett.122.170403} {\bibfield  {journal} {\bibinfo  {journal}
  {Phys. Rev. Lett.}\ }\textbf {\bibinfo {volume} {122}},\ \bibinfo {pages}
  {170403} (\bibinfo {year} {2019})}\BibitemShut {NoStop}%
\bibitem [{\citenamefont {Balachandran}\ \emph {et~al.}(2019)\citenamefont
  {Balachandran}, \citenamefont {Clark}, \citenamefont {Goold},\ and\
  \citenamefont {Poletti}}]{device-1}%
  \BibitemOpen
  \bibfield  {author} {\bibinfo {author} {\bibfnamefont {V.}~\bibnamefont
  {Balachandran}}, \bibinfo {author} {\bibfnamefont {S.~R.}\ \bibnamefont
  {Clark}}, \bibinfo {author} {\bibfnamefont {J.}~\bibnamefont {Goold}}, \ and\
  \bibinfo {author} {\bibfnamefont {D.}~\bibnamefont {Poletti}},\ }\href
  {\doibase 10.1103/PhysRevLett.123.020603} {\bibfield  {journal} {\bibinfo
  {journal} {Phys. Rev. Lett.}\ }\textbf {\bibinfo {volume} {123}},\ \bibinfo
  {pages} {020603} (\bibinfo {year} {2019})}\BibitemShut {NoStop}%
\bibitem [{\citenamefont {Chiaracane}\ \emph {et~al.}(2020)\citenamefont
  {Chiaracane}, \citenamefont {Mitchison}, \citenamefont {Purkayastha},
  \citenamefont {Haack},\ and\ \citenamefont {Goold}}]{device-2}%
  \BibitemOpen
  \bibfield  {author} {\bibinfo {author} {\bibfnamefont {C.}~\bibnamefont
  {Chiaracane}}, \bibinfo {author} {\bibfnamefont {M.~T.}\ \bibnamefont
  {Mitchison}}, \bibinfo {author} {\bibfnamefont {A.}~\bibnamefont
  {Purkayastha}}, \bibinfo {author} {\bibfnamefont {G.}~\bibnamefont {Haack}},
  \ and\ \bibinfo {author} {\bibfnamefont {J.}~\bibnamefont {Goold}},\ }\href
  {\doibase 10.1103/PhysRevResearch.2.013093} {\bibfield  {journal} {\bibinfo
  {journal} {Phys. Rev. Research}\ }\textbf {\bibinfo {volume} {2}},\ \bibinfo
  {pages} {013093} (\bibinfo {year} {2020})}\BibitemShut {NoStop}%
\bibitem [{\citenamefont {Saha}\ and\ \citenamefont
  {Maiti}(2019)}]{quasi-periodic_rectifier}%
  \BibitemOpen
  \bibfield  {author} {\bibinfo {author} {\bibfnamefont {M.}~\bibnamefont
  {Saha}}\ and\ \bibinfo {author} {\bibfnamefont {S.~K.}\ \bibnamefont
  {Maiti}},\ }\href {\doibase 10.1088/1361-6463/ab3a0e} {\bibfield  {journal}
  {\bibinfo  {journal} {Journal of Physics D: Applied Physics}\ }\textbf
  {\bibinfo {volume} {52}},\ \bibinfo {pages} {465304} (\bibinfo {year}
  {2019})}\BibitemShut {NoStop}%
\bibitem [{\citenamefont {Saha}\ and\ \citenamefont
  {Maiti}(2017)}]{quasi-periodic-rectifier3}%
  \BibitemOpen
  \bibfield  {author} {\bibinfo {author} {\bibfnamefont {M.}~\bibnamefont
  {Saha}}\ and\ \bibinfo {author} {\bibfnamefont {S.~K.}\ \bibnamefont
  {Maiti}},\ }\href {\doibase https://doi.org/10.1016/j.physe.2017.06.023}
  {\bibfield  {journal} {\bibinfo  {journal} {Physica E: Low-dimensional
  Systems and Nanostructures}\ }\textbf {\bibinfo {volume} {93}},\ \bibinfo
  {pages} {275} (\bibinfo {year} {2017})}\BibitemShut {NoStop}%
\bibitem [{\citenamefont {Ostlund}\ \emph
  {et~al.}(1983{\natexlab{b}})\citenamefont {Ostlund}, \citenamefont {Pandit},
  \citenamefont {Rand}, \citenamefont {Schellnhuber},\ and\ \citenamefont
  {Siggia}}]{Fibonacci-first}%
  \BibitemOpen
  \bibfield  {author} {\bibinfo {author} {\bibfnamefont {S.}~\bibnamefont
  {Ostlund}}, \bibinfo {author} {\bibfnamefont {R.}~\bibnamefont {Pandit}},
  \bibinfo {author} {\bibfnamefont {D.}~\bibnamefont {Rand}}, \bibinfo {author}
  {\bibfnamefont {H.~J.}\ \bibnamefont {Schellnhuber}}, \ and\ \bibinfo
  {author} {\bibfnamefont {E.~D.}\ \bibnamefont {Siggia}},\ }\href {\doibase
  10.1103/PhysRevLett.50.1873} {\bibfield  {journal} {\bibinfo  {journal}
  {Phys. Rev. Lett.}\ }\textbf {\bibinfo {volume} {50}},\ \bibinfo {pages}
  {1873} (\bibinfo {year} {1983}{\natexlab{b}})}\BibitemShut {NoStop}%
\bibitem [{\citenamefont {Kohmoto}\ \emph {et~al.}(1987)\citenamefont
  {Kohmoto}, \citenamefont {Sutherland},\ and\ \citenamefont
  {Tang}}]{Fibonacci-second}%
  \BibitemOpen
  \bibfield  {author} {\bibinfo {author} {\bibfnamefont {M.}~\bibnamefont
  {Kohmoto}}, \bibinfo {author} {\bibfnamefont {B.}~\bibnamefont {Sutherland}},
  \ and\ \bibinfo {author} {\bibfnamefont {C.}~\bibnamefont {Tang}},\ }\href
  {\doibase 10.1103/PhysRevB.35.1020} {\bibfield  {journal} {\bibinfo
  {journal} {Phys. Rev. B}\ }\textbf {\bibinfo {volume} {35}},\ \bibinfo
  {pages} {1020} (\bibinfo {year} {1987})}\BibitemShut {NoStop}%
\bibitem [{\citenamefont {Purkayastha}\ \emph {et~al.}(2017)\citenamefont
  {Purkayastha}, \citenamefont {Dhar},\ and\ \citenamefont
  {Kulkarni}}]{Archak_phase_diagram}%
  \BibitemOpen
  \bibfield  {author} {\bibinfo {author} {\bibfnamefont {A.}~\bibnamefont
  {Purkayastha}}, \bibinfo {author} {\bibfnamefont {A.}~\bibnamefont {Dhar}}, \
  and\ \bibinfo {author} {\bibfnamefont {M.}~\bibnamefont {Kulkarni}},\ }\href
  {\doibase 10.1103/PhysRevB.96.180204} {\bibfield  {journal} {\bibinfo
  {journal} {Phys. Rev. B}\ }\textbf {\bibinfo {volume} {96}},\ \bibinfo
  {pages} {180204} (\bibinfo {year} {2017})}\BibitemShut {NoStop}%
\bibitem [{\citenamefont {Purkayastha}\ \emph {et~al.}(2018)\citenamefont
  {Purkayastha}, \citenamefont {Sanyal}, \citenamefont {Dhar},\ and\
  \citenamefont {Kulkarni}}]{Archak_AAH}%
  \BibitemOpen
  \bibfield  {author} {\bibinfo {author} {\bibfnamefont {A.}~\bibnamefont
  {Purkayastha}}, \bibinfo {author} {\bibfnamefont {S.}~\bibnamefont {Sanyal}},
  \bibinfo {author} {\bibfnamefont {A.}~\bibnamefont {Dhar}}, \ and\ \bibinfo
  {author} {\bibfnamefont {M.}~\bibnamefont {Kulkarni}},\ }\href {\doibase
  10.1103/PhysRevB.97.174206} {\bibfield  {journal} {\bibinfo  {journal} {Phys.
  Rev. B}\ }\textbf {\bibinfo {volume} {97}},\ \bibinfo {pages} {174206}
  (\bibinfo {year} {2018})}\BibitemShut {NoStop}%
\bibitem [{\citenamefont {Saha}\ \emph {et~al.}(2019)\citenamefont {Saha},
  \citenamefont {Maiti},\ and\ \citenamefont {Purkayastha}}]{Madhumita_GAAH}%
  \BibitemOpen
  \bibfield  {author} {\bibinfo {author} {\bibfnamefont {M.}~\bibnamefont
  {Saha}}, \bibinfo {author} {\bibfnamefont {S.~K.}\ \bibnamefont {Maiti}}, \
  and\ \bibinfo {author} {\bibfnamefont {A.}~\bibnamefont {Purkayastha}},\
  }\href {\doibase 10.1103/PhysRevB.100.174201} {\bibfield  {journal} {\bibinfo
   {journal} {Phys. Rev. B}\ }\textbf {\bibinfo {volume} {100}},\ \bibinfo
  {pages} {174201} (\bibinfo {year} {2019})}\BibitemShut {NoStop}%
\bibitem [{\citenamefont {Saha}\ \emph {et~al.}(2021)\citenamefont {Saha},
  \citenamefont {Agarwalla},\ and\ \citenamefont
  {Venkatesh}}]{Madhumita_readout}%
  \BibitemOpen
  \bibfield  {author} {\bibinfo {author} {\bibfnamefont {M.}~\bibnamefont
  {Saha}}, \bibinfo {author} {\bibfnamefont {B.~K.}\ \bibnamefont {Agarwalla}},
  \ and\ \bibinfo {author} {\bibfnamefont {B.~P.}\ \bibnamefont {Venkatesh}},\
  }\href {\doibase 10.1103/PhysRevA.103.023330} {\bibfield  {journal} {\bibinfo
   {journal} {Phys. Rev. A}\ }\textbf {\bibinfo {volume} {103}},\ \bibinfo
  {pages} {023330} (\bibinfo {year} {2021})}\BibitemShut {NoStop}%
\bibitem [{\citenamefont {Varma}\ \emph {et~al.}(2017)\citenamefont {Varma},
  \citenamefont {de~Mulatier},\ and\ \citenamefont {\ifmmode \check{Z}\else
  \v{Z}\fi{}nidari\ifmmode~\check{c}\else \v{c}\fi{}}}]{Varma_quasi_periodic}%
  \BibitemOpen
  \bibfield  {author} {\bibinfo {author} {\bibfnamefont {V.~K.}\ \bibnamefont
  {Varma}}, \bibinfo {author} {\bibfnamefont {C.}~\bibnamefont {de~Mulatier}},
  \ and\ \bibinfo {author} {\bibfnamefont {M.}~\bibnamefont {\ifmmode
  \check{Z}\else \v{Z}\fi{}nidari\ifmmode~\check{c}\else \v{c}\fi{}}},\ }\href
  {\doibase 10.1103/PhysRevE.96.032130} {\bibfield  {journal} {\bibinfo
  {journal} {Phys. Rev. E}\ }\textbf {\bibinfo {volume} {96}},\ \bibinfo
  {pages} {032130} (\bibinfo {year} {2017})}\BibitemShut {NoStop}%
\bibitem [{\citenamefont {{\v Z}nidari{\v c}}\ and\ \citenamefont
  {Ljubotina}(2018)}]{interaction_quasi_periodic}%
  \BibitemOpen
  \bibfield  {author} {\bibinfo {author} {\bibfnamefont {M.}~\bibnamefont {{\v
  Z}nidari{\v c}}}\ and\ \bibinfo {author} {\bibfnamefont {M.}~\bibnamefont
  {Ljubotina}},\ }\href {\doibase 10.1073/pnas.1800589115} {\bibfield
  {journal} {\bibinfo  {journal} {Proceedings of the National Academy of
  Sciences}\ }\textbf {\bibinfo {volume} {115}},\ \bibinfo {pages} {4595}
  (\bibinfo {year} {2018})}\BibitemShut {NoStop}%
\bibitem [{\citenamefont {Varma}\ and\ \citenamefont {\ifmmode \check{Z}\else
  \v{Z}\fi{}nidari\ifmmode~\check{c}\else
  \v{c}\fi{}}(2019)}]{interaction_Fibonacci}%
  \BibitemOpen
  \bibfield  {author} {\bibinfo {author} {\bibfnamefont {V.~K.}\ \bibnamefont
  {Varma}}\ and\ \bibinfo {author} {\bibfnamefont {M.}~\bibnamefont {\ifmmode
  \check{Z}\else \v{Z}\fi{}nidari\ifmmode~\check{c}\else \v{c}\fi{}}},\ }\href
  {\doibase 10.1103/PhysRevB.100.085105} {\bibfield  {journal} {\bibinfo
  {journal} {Phys. Rev. B}\ }\textbf {\bibinfo {volume} {100}},\ \bibinfo
  {pages} {085105} (\bibinfo {year} {2019})}\BibitemShut {NoStop}%
\bibitem [{\citenamefont {{\v{Z}}nidari{\v{c}}}\ and\ \citenamefont
  {Horvat}(2013)}]{dephasing_disorder}%
  \BibitemOpen
  \bibfield  {author} {\bibinfo {author} {\bibfnamefont {M.}~\bibnamefont
  {{\v{Z}}nidari{\v{c}}}}\ and\ \bibinfo {author} {\bibfnamefont
  {M.}~\bibnamefont {Horvat}},\ }\href {\doibase 10.1140/epjb/e2012-30730-9}
  {\bibfield  {journal} {\bibinfo  {journal} {The European Physical Journal B}\
  }\textbf {\bibinfo {volume} {86}},\ \bibinfo {pages} {67} (\bibinfo {year}
  {2013})}\BibitemShut {NoStop}%
\bibitem [{\citenamefont {Medvedyeva}\ \emph {et~al.}(2016)\citenamefont
  {Medvedyeva}, \citenamefont {Prosen},\ and\ \citenamefont {\ifmmode
  \check{Z}\else \v{Z}\fi{}nidari\ifmmode~\check{c}\else
  \v{c}\fi{}}}]{dephasing-interaction1}%
  \BibitemOpen
  \bibfield  {author} {\bibinfo {author} {\bibfnamefont {M.~V.}\ \bibnamefont
  {Medvedyeva}}, \bibinfo {author} {\bibfnamefont {T.~c.~v.}\ \bibnamefont
  {Prosen}}, \ and\ \bibinfo {author} {\bibfnamefont {M.}~\bibnamefont
  {\ifmmode \check{Z}\else \v{Z}\fi{}nidari\ifmmode~\check{c}\else
  \v{c}\fi{}}},\ }\href {\doibase 10.1103/PhysRevB.93.094205} {\bibfield
  {journal} {\bibinfo  {journal} {Phys. Rev. B}\ }\textbf {\bibinfo {volume}
  {93}},\ \bibinfo {pages} {094205} (\bibinfo {year} {2016})}\BibitemShut
  {NoStop}%
\bibitem [{\citenamefont {Žnidarič}\ \emph {et~al.}(2017)\citenamefont
  {Žnidarič}, \citenamefont {Mendoza-Arenas}, \citenamefont {Clark},\ and\
  \citenamefont {Goold}}]{dephasing_spin_interaction2}%
  \BibitemOpen
  \bibfield  {author} {\bibinfo {author} {\bibfnamefont {M.}~\bibnamefont
  {Žnidarič}}, \bibinfo {author} {\bibfnamefont {J.~J.}\ \bibnamefont
  {Mendoza-Arenas}}, \bibinfo {author} {\bibfnamefont {S.~R.}\ \bibnamefont
  {Clark}}, \ and\ \bibinfo {author} {\bibfnamefont {J.}~\bibnamefont
  {Goold}},\ }\href {\doibase https://doi.org/10.1002/andp.201600298}
  {\bibfield  {journal} {\bibinfo  {journal} {Annalen der Physik}\ }\textbf
  {\bibinfo {volume} {529}},\ \bibinfo {pages} {1600298} (\bibinfo {year}
  {2017})}\BibitemShut {NoStop}%
\bibitem [{\citenamefont {\ifmmode \check{Z}\else
  \v{Z}\fi{}nidari\ifmmode~\check{c}\else
  \v{c}\fi{}}(2018)}]{dephasing_interaction3}%
  \BibitemOpen
  \bibfield  {author} {\bibinfo {author} {\bibfnamefont {M.}~\bibnamefont
  {\ifmmode \check{Z}\else \v{Z}\fi{}nidari\ifmmode~\check{c}\else
  \v{c}\fi{}}},\ }\href {\doibase 10.1103/PhysRevB.97.214202} {\bibfield
  {journal} {\bibinfo  {journal} {Phys. Rev. B}\ }\textbf {\bibinfo {volume}
  {97}},\ \bibinfo {pages} {214202} (\bibinfo {year} {2018})}\BibitemShut
  {NoStop}%
\bibitem [{\citenamefont {Lacerda}\ \emph {et~al.}(2021)\citenamefont
  {Lacerda}, \citenamefont {Goold},\ and\ \citenamefont
  {Landi}}]{Landi-latest}%
  \BibitemOpen
  \bibfield  {author} {\bibinfo {author} {\bibfnamefont {A.~M.}\ \bibnamefont
  {Lacerda}}, \bibinfo {author} {\bibfnamefont {J.}~\bibnamefont {Goold}}, \
  and\ \bibinfo {author} {\bibfnamefont {G.~T.}\ \bibnamefont {Landi}},\ }\href
  {\doibase 10.1103/PhysRevB.104.174203} {\bibfield  {journal} {\bibinfo
  {journal} {Phys. Rev. B}\ }\textbf {\bibinfo {volume} {104}},\ \bibinfo
  {pages} {174203} (\bibinfo {year} {2021})}\BibitemShut {NoStop}%
\bibitem [{\citenamefont {Dwiputra}\ and\ \citenamefont {Zen}(2021)}]{ENQT1}%
  \BibitemOpen
  \bibfield  {author} {\bibinfo {author} {\bibfnamefont {D.}~\bibnamefont
  {Dwiputra}}\ and\ \bibinfo {author} {\bibfnamefont {F.~P.}\ \bibnamefont
  {Zen}},\ }\href {\doibase 10.1103/PhysRevA.104.022205} {\bibfield  {journal}
  {\bibinfo  {journal} {Phys. Rev. A}\ }\textbf {\bibinfo {volume} {104}},\
  \bibinfo {pages} {022205} (\bibinfo {year} {2021})}\BibitemShut {NoStop}%
\bibitem [{\citenamefont {Chiaracane}\ \emph {et~al.}(2021)\citenamefont
  {Chiaracane}, \citenamefont {Purkayastha}, \citenamefont {Mitchison},\ and\
  \citenamefont {Goold}}]{buttikerdephase}%
  \BibitemOpen
  \bibfield  {author} {\bibinfo {author} {\bibfnamefont {C.}~\bibnamefont
  {Chiaracane}}, \bibinfo {author} {\bibfnamefont {A.}~\bibnamefont
  {Purkayastha}}, \bibinfo {author} {\bibfnamefont {M.~T.}\ \bibnamefont
  {Mitchison}}, \ and\ \bibinfo {author} {\bibfnamefont {J.}~\bibnamefont
  {Goold}},\ }\href@noop {} {\bibfield  {journal} {\bibinfo  {journal} {arXiv
  preprint arXiv:2112.02035}\ } (\bibinfo {year} {2021})}\BibitemShut {NoStop}%
\bibitem [{\citenamefont {Zerah-Harush}\ and\ \citenamefont
  {Dubi}(2020)}]{dephasing_dubi}%
  \BibitemOpen
  \bibfield  {author} {\bibinfo {author} {\bibfnamefont {E.}~\bibnamefont
  {Zerah-Harush}}\ and\ \bibinfo {author} {\bibfnamefont {Y.}~\bibnamefont
  {Dubi}},\ }\href {\doibase 10.1103/PhysRevResearch.2.023294} {\bibfield
  {journal} {\bibinfo  {journal} {Phys. Rev. Research}\ }\textbf {\bibinfo
  {volume} {2}},\ \bibinfo {pages} {023294} (\bibinfo {year}
  {2020})}\BibitemShut {NoStop}%
\bibitem [{\citenamefont {Cattena}\ \emph {et~al.}(2010)\citenamefont
  {Cattena}, \citenamefont {Bustos-Mar\'un},\ and\ \citenamefont
  {Pastawski}}]{molecular_dephasing1}%
  \BibitemOpen
  \bibfield  {author} {\bibinfo {author} {\bibfnamefont {C.~J.}\ \bibnamefont
  {Cattena}}, \bibinfo {author} {\bibfnamefont {R.~A.}\ \bibnamefont
  {Bustos-Mar\'un}}, \ and\ \bibinfo {author} {\bibfnamefont {H.~M.}\
  \bibnamefont {Pastawski}},\ }\href {\doibase 10.1103/PhysRevB.82.144201}
  {\bibfield  {journal} {\bibinfo  {journal} {Phys. Rev. B}\ }\textbf {\bibinfo
  {volume} {82}},\ \bibinfo {pages} {144201} (\bibinfo {year}
  {2010})}\BibitemShut {NoStop}%
\bibitem [{\citenamefont {Nozaki}\ \emph {et~al.}(2012)\citenamefont {Nozaki},
  \citenamefont {Gomes~da Rocha}, \citenamefont {Pastawski},\ and\
  \citenamefont {Cuniberti}}]{molecular_dephasing2}%
  \BibitemOpen
  \bibfield  {author} {\bibinfo {author} {\bibfnamefont {D.}~\bibnamefont
  {Nozaki}}, \bibinfo {author} {\bibfnamefont {C.}~\bibnamefont {Gomes~da
  Rocha}}, \bibinfo {author} {\bibfnamefont {H.~M.}\ \bibnamefont {Pastawski}},
  \ and\ \bibinfo {author} {\bibfnamefont {G.}~\bibnamefont {Cuniberti}},\
  }\href {\doibase 10.1103/PhysRevB.85.155327} {\bibfield  {journal} {\bibinfo
  {journal} {Phys. Rev. B}\ }\textbf {\bibinfo {volume} {85}},\ \bibinfo
  {pages} {155327} (\bibinfo {year} {2012})}\BibitemShut {NoStop}%
\bibitem [{\citenamefont {Maassen}\ \emph {et~al.}(2009)\citenamefont
  {Maassen}, \citenamefont {Zahid},\ and\ \citenamefont
  {Guo}}]{molecular_dephasing3}%
  \BibitemOpen
  \bibfield  {author} {\bibinfo {author} {\bibfnamefont {J.}~\bibnamefont
  {Maassen}}, \bibinfo {author} {\bibfnamefont {F.}~\bibnamefont {Zahid}}, \
  and\ \bibinfo {author} {\bibfnamefont {H.}~\bibnamefont {Guo}},\ }\href
  {\doibase 10.1103/PhysRevB.80.125423} {\bibfield  {journal} {\bibinfo
  {journal} {Phys. Rev. B}\ }\textbf {\bibinfo {volume} {80}},\ \bibinfo
  {pages} {125423} (\bibinfo {year} {2009})}\BibitemShut {NoStop}%
\bibitem [{\citenamefont {Nozaki}\ \emph {et~al.}(2008)\citenamefont {Nozaki},
  \citenamefont {Girard},\ and\ \citenamefont
  {Yoshizawa}}]{molecular_dephasing4}%
  \BibitemOpen
  \bibfield  {author} {\bibinfo {author} {\bibfnamefont {D.}~\bibnamefont
  {Nozaki}}, \bibinfo {author} {\bibfnamefont {Y.}~\bibnamefont {Girard}}, \
  and\ \bibinfo {author} {\bibfnamefont {K.}~\bibnamefont {Yoshizawa}},\ }\href
  {\doibase 10.1021/jp806806j} {\bibfield  {journal} {\bibinfo  {journal} {The
  Journal of Physical Chemistry C}\ }\textbf {\bibinfo {volume} {112}},\
  \bibinfo {pages} {17408} (\bibinfo {year} {2008})}\BibitemShut {NoStop}%
\bibitem [{\citenamefont {Kilgour}\ and\ \citenamefont
  {Segal}(2016)}]{segal-probe-1}%
  \BibitemOpen
  \bibfield  {author} {\bibinfo {author} {\bibfnamefont {M.}~\bibnamefont
  {Kilgour}}\ and\ \bibinfo {author} {\bibfnamefont {D.}~\bibnamefont
  {Segal}},\ }\href {\doibase 10.1063/1.4944470} {\bibfield  {journal}
  {\bibinfo  {journal} {The Journal of Chemical Physics}\ }\textbf {\bibinfo
  {volume} {144}},\ \bibinfo {pages} {124107} (\bibinfo {year}
  {2016})}\BibitemShut {NoStop}%
\bibitem [{\citenamefont {Kilgour}\ and\ \citenamefont
  {Segal}(2015)}]{segal-probe-2}%
  \BibitemOpen
  \bibfield  {author} {\bibinfo {author} {\bibfnamefont {M.}~\bibnamefont
  {Kilgour}}\ and\ \bibinfo {author} {\bibfnamefont {D.}~\bibnamefont
  {Segal}},\ }\href {\doibase 10.1063/1.4926395} {\bibfield  {journal}
  {\bibinfo  {journal} {The Journal of Chemical Physics}\ }\textbf {\bibinfo
  {volume} {143}},\ \bibinfo {pages} {024111} (\bibinfo {year}
  {2015})}\BibitemShut {NoStop}%
\bibitem [{\citenamefont {Korol}\ \emph {et~al.}(2018)\citenamefont {Korol},
  \citenamefont {Kilgour},\ and\ \citenamefont {Segal}}]{segal-probe-3}%
  \BibitemOpen
  \bibfield  {author} {\bibinfo {author} {\bibfnamefont {R.}~\bibnamefont
  {Korol}}, \bibinfo {author} {\bibfnamefont {M.}~\bibnamefont {Kilgour}}, \
  and\ \bibinfo {author} {\bibfnamefont {D.}~\bibnamefont {Segal}},\ }\href
  {\doibase https://doi.org/10.1016/j.cpc.2017.10.005} {\bibfield  {journal}
  {\bibinfo  {journal} {Computer Physics Communications}\ }\textbf {\bibinfo
  {volume} {224}},\ \bibinfo {pages} {396} (\bibinfo {year}
  {2018})}\BibitemShut {NoStop}%
\bibitem [{\citenamefont {Bandyopadhyay}\ and\ \citenamefont
  {Segal}(2011)}]{malay}%
  \BibitemOpen
  \bibfield  {author} {\bibinfo {author} {\bibfnamefont {M.}~\bibnamefont
  {Bandyopadhyay}}\ and\ \bibinfo {author} {\bibfnamefont {D.}~\bibnamefont
  {Segal}},\ }\href {\doibase 10.1103/PhysRevE.84.011151} {\bibfield  {journal}
  {\bibinfo  {journal} {Phys. Rev. E}\ }\textbf {\bibinfo {volume} {84}},\
  \bibinfo {pages} {011151} (\bibinfo {year} {2011})}\BibitemShut {NoStop}%
\bibitem [{\citenamefont {Bedkihal}\ \emph {et~al.}(2013)\citenamefont
  {Bedkihal}, \citenamefont {Bandyopadhyay},\ and\ \citenamefont
  {Segal}}]{Bedkihal2013}%
  \BibitemOpen
  \bibfield  {author} {\bibinfo {author} {\bibfnamefont {S.}~\bibnamefont
  {Bedkihal}}, \bibinfo {author} {\bibfnamefont {M.}~\bibnamefont
  {Bandyopadhyay}}, \ and\ \bibinfo {author} {\bibfnamefont {D.}~\bibnamefont
  {Segal}},\ }\href {\doibase 10.1140/epjb/e2013-40971-7} {\bibfield  {journal}
  {\bibinfo  {journal} {The European Physical Journal B}\ }\textbf {\bibinfo
  {volume} {86}},\ \bibinfo {pages} {506} (\bibinfo {year} {2013})}\BibitemShut
  {NoStop}%
\bibitem [{\citenamefont {D'Amato}\ and\ \citenamefont
  {Pastawski}(1990)}]{Pastawski}%
  \BibitemOpen
  \bibfield  {author} {\bibinfo {author} {\bibfnamefont {J.~L.}\ \bibnamefont
  {D'Amato}}\ and\ \bibinfo {author} {\bibfnamefont {H.~M.}\ \bibnamefont
  {Pastawski}},\ }\href {\doibase 10.1103/PhysRevB.41.7411} {\bibfield
  {journal} {\bibinfo  {journal} {Phys. Rev. B}\ }\textbf {\bibinfo {volume}
  {41}},\ \bibinfo {pages} {7411} (\bibinfo {year} {1990})}\BibitemShut
  {NoStop}%
\bibitem [{\citenamefont {Roy}(2007)}]{Roy-1}%
  \BibitemOpen
  \bibfield  {author} {\bibinfo {author} {\bibfnamefont {D.}~\bibnamefont
  {Roy}},\ }\href {\doibase 10.1088/0953-8984/20/02/025206} {\bibfield
  {journal} {\bibinfo  {journal} {Journal of Physics: Condensed Matter}\
  }\textbf {\bibinfo {volume} {20}},\ \bibinfo {pages} {025206} (\bibinfo
  {year} {2007})}\BibitemShut {NoStop}%
\bibitem [{\citenamefont {Roy}\ and\ \citenamefont
  {Dhar}(2007)}]{self-2-Abhishek}%
  \BibitemOpen
  \bibfield  {author} {\bibinfo {author} {\bibfnamefont {D.}~\bibnamefont
  {Roy}}\ and\ \bibinfo {author} {\bibfnamefont {A.}~\bibnamefont {Dhar}},\
  }\href {\doibase 10.1103/PhysRevB.75.195110} {\bibfield  {journal} {\bibinfo
  {journal} {Phys. Rev. B}\ }\textbf {\bibinfo {volume} {75}},\ \bibinfo
  {pages} {195110} (\bibinfo {year} {2007})}\BibitemShut {NoStop}%
\bibitem [{\citenamefont {B\"uttiker}(1985)}]{Bu-1}%
  \BibitemOpen
  \bibfield  {author} {\bibinfo {author} {\bibfnamefont {M.}~\bibnamefont
  {B\"uttiker}},\ }\href {\doibase 10.1103/PhysRevB.32.1846} {\bibfield
  {journal} {\bibinfo  {journal} {Phys. Rev. B}\ }\textbf {\bibinfo {volume}
  {32}},\ \bibinfo {pages} {1846} (\bibinfo {year} {1985})}\BibitemShut
  {NoStop}%
\bibitem [{\citenamefont {B\"uttiker}(1986)}]{Bu-2}%
  \BibitemOpen
  \bibfield  {author} {\bibinfo {author} {\bibfnamefont {M.}~\bibnamefont
  {B\"uttiker}},\ }\href {\doibase 10.1103/PhysRevB.33.3020} {\bibfield
  {journal} {\bibinfo  {journal} {Phys. Rev. B}\ }\textbf {\bibinfo {volume}
  {33}},\ \bibinfo {pages} {3020} (\bibinfo {year} {1986})}\BibitemShut
  {NoStop}%
\bibitem [{\citenamefont {Pilgram}\ \emph {et~al.}(2006)\citenamefont
  {Pilgram}, \citenamefont {Samuelsson}, \citenamefont {F\"orster},\ and\
  \citenamefont {B\"uttiker}}]{Bu-3}%
  \BibitemOpen
  \bibfield  {author} {\bibinfo {author} {\bibfnamefont {S.}~\bibnamefont
  {Pilgram}}, \bibinfo {author} {\bibfnamefont {P.}~\bibnamefont {Samuelsson}},
  \bibinfo {author} {\bibfnamefont {H.}~\bibnamefont {F\"orster}}, \ and\
  \bibinfo {author} {\bibfnamefont {M.}~\bibnamefont {B\"uttiker}},\ }\href
  {\doibase 10.1103/PhysRevLett.97.066801} {\bibfield  {journal} {\bibinfo
  {journal} {Phys. Rev. Lett.}\ }\textbf {\bibinfo {volume} {97}},\ \bibinfo
  {pages} {066801} (\bibinfo {year} {2006})}\BibitemShut {NoStop}%
\bibitem [{\citenamefont {F\"orster}\ \emph {et~al.}(2007)\citenamefont
  {F\"orster}, \citenamefont {Samuelsson}, \citenamefont {Pilgram},\ and\
  \citenamefont {B\"uttiker}}]{Bu-4}%
  \BibitemOpen
  \bibfield  {author} {\bibinfo {author} {\bibfnamefont {H.}~\bibnamefont
  {F\"orster}}, \bibinfo {author} {\bibfnamefont {P.}~\bibnamefont
  {Samuelsson}}, \bibinfo {author} {\bibfnamefont {S.}~\bibnamefont {Pilgram}},
  \ and\ \bibinfo {author} {\bibfnamefont {M.}~\bibnamefont {B\"uttiker}},\
  }\href {\doibase 10.1103/PhysRevB.75.035340} {\bibfield  {journal} {\bibinfo
  {journal} {Phys. Rev. B}\ }\textbf {\bibinfo {volume} {75}},\ \bibinfo
  {pages} {035340} (\bibinfo {year} {2007})}\BibitemShut {NoStop}%
\bibitem [{\citenamefont {Saryal}\ \emph {et~al.}(2021)\citenamefont {Saryal},
  \citenamefont {Gerry}, \citenamefont {Khait}, \citenamefont {Segal},\ and\
  \citenamefont {Agarwalla}}]{Saryal-bound}%
  \BibitemOpen
  \bibfield  {author} {\bibinfo {author} {\bibfnamefont {S.}~\bibnamefont
  {Saryal}}, \bibinfo {author} {\bibfnamefont {M.}~\bibnamefont {Gerry}},
  \bibinfo {author} {\bibfnamefont {I.}~\bibnamefont {Khait}}, \bibinfo
  {author} {\bibfnamefont {D.}~\bibnamefont {Segal}}, \ and\ \bibinfo {author}
  {\bibfnamefont {B.~K.}\ \bibnamefont {Agarwalla}},\ }\href {\doibase
  10.1103/PhysRevLett.127.190603} {\bibfield  {journal} {\bibinfo  {journal}
  {Phys. Rev. Lett.}\ }\textbf {\bibinfo {volume} {127}},\ \bibinfo {pages}
  {190603} (\bibinfo {year} {2021})}\BibitemShut {NoStop}%
\bibitem [{\citenamefont {Pietzonka}\ and\ \citenamefont
  {Seifert}(2018)}]{Udo-TUR1}%
  \BibitemOpen
  \bibfield  {author} {\bibinfo {author} {\bibfnamefont {P.}~\bibnamefont
  {Pietzonka}}\ and\ \bibinfo {author} {\bibfnamefont {U.}~\bibnamefont
  {Seifert}},\ }\href {\doibase 10.1103/PhysRevLett.120.190602} {\bibfield
  {journal} {\bibinfo  {journal} {Phys. Rev. Lett.}\ }\textbf {\bibinfo
  {volume} {120}},\ \bibinfo {pages} {190602} (\bibinfo {year}
  {2018})}\BibitemShut {NoStop}%
\bibitem [{\citenamefont {Schwinger}(1961)}]{Keldysh1}%
  \BibitemOpen
  \bibfield  {author} {\bibinfo {author} {\bibfnamefont {J.}~\bibnamefont
  {Schwinger}},\ }\href {\doibase 10.1063/1.1703727} {\bibfield  {journal}
  {\bibinfo  {journal} {Journal of Mathematical Physics}\ }\textbf {\bibinfo
  {volume} {2}},\ \bibinfo {pages} {407} (\bibinfo {year} {1961})}\BibitemShut
  {NoStop}%
\bibitem [{\citenamefont {Keldysh}(1964)}]{Keldysh2}%
  \BibitemOpen
  \bibfield  {author} {\bibinfo {author} {\bibfnamefont {L.~V.}\ \bibnamefont
  {Keldysh}},\ }\href@noop {} {\emph {\bibinfo {title} {Diagram technique for
  nonequilibrium processes}}}\ (\bibinfo {year} {1964})\BibitemShut {NoStop}%
\bibitem [{\citenamefont {Rammer}\ and\ \citenamefont
  {Smith}(1986)}]{Keldysh3}%
  \BibitemOpen
  \bibfield  {author} {\bibinfo {author} {\bibfnamefont {J.}~\bibnamefont
  {Rammer}}\ and\ \bibinfo {author} {\bibfnamefont {H.}~\bibnamefont {Smith}},\
  }\href {\doibase 10.1103/RevModPhys.58.323} {\bibfield  {journal} {\bibinfo
  {journal} {Rev. Mod. Phys.}\ }\textbf {\bibinfo {volume} {58}},\ \bibinfo
  {pages} {323} (\bibinfo {year} {1986})}\BibitemShut {NoStop}%
\bibitem [{\citenamefont {Haug}\ and\ \citenamefont {Jauho}(2008)}]{Keldysh4}%
  \BibitemOpen
  \bibfield  {author} {\bibinfo {author} {\bibfnamefont {H.}~\bibnamefont
  {Haug}}\ and\ \bibinfo {author} {\bibfnamefont {A.}~\bibnamefont {Jauho}},\
  }\href@noop {} {\emph {\bibinfo {title} {Quantum Kinetics in Transport and
  Optics of Semiconductors}}}\ (\bibinfo {year} {Springer, New York,
  2008})\BibitemShut {NoStop}%
\bibitem [{\citenamefont {Wang}\ \emph {et~al.}(2014)\citenamefont {Wang},
  \citenamefont {Agarwalla}, \citenamefont {Li},\ and\ \citenamefont
  {Thingna}}]{Wang2014}%
  \BibitemOpen
  \bibfield  {author} {\bibinfo {author} {\bibfnamefont {J.-S.}\ \bibnamefont
  {Wang}}, \bibinfo {author} {\bibfnamefont {B.~K.}\ \bibnamefont {Agarwalla}},
  \bibinfo {author} {\bibfnamefont {H.}~\bibnamefont {Li}}, \ and\ \bibinfo
  {author} {\bibfnamefont {J.}~\bibnamefont {Thingna}},\ }\href {\doibase
  10.1007/s11467-013-0340-x} {\bibfield  {journal} {\bibinfo  {journal}
  {Frontiers of Physics}\ }\textbf {\bibinfo {volume} {9}},\ \bibinfo {pages}
  {673} (\bibinfo {year} {2014})}\BibitemShut {NoStop}%
\bibitem [{\citenamefont {Meir}\ and\ \citenamefont
  {Wingreen}(1992)}]{Wingreen}%
  \BibitemOpen
  \bibfield  {author} {\bibinfo {author} {\bibfnamefont {Y.}~\bibnamefont
  {Meir}}\ and\ \bibinfo {author} {\bibfnamefont {N.~S.}\ \bibnamefont
  {Wingreen}},\ }\href {\doibase 10.1103/PhysRevLett.68.2512} {\bibfield
  {journal} {\bibinfo  {journal} {Phys. Rev. Lett.}\ }\textbf {\bibinfo
  {volume} {68}},\ \bibinfo {pages} {2512} (\bibinfo {year}
  {1992})}\BibitemShut {NoStop}%
\bibitem [{\citenamefont {Dhar}\ and\ \citenamefont {Sen}(2006)}]{bound-NEGF}%
  \BibitemOpen
  \bibfield  {author} {\bibinfo {author} {\bibfnamefont {A.}~\bibnamefont
  {Dhar}}\ and\ \bibinfo {author} {\bibfnamefont {D.}~\bibnamefont {Sen}},\
  }\href {\doibase 10.1103/PhysRevB.73.085119} {\bibfield  {journal} {\bibinfo
  {journal} {Phys. Rev. B}\ }\textbf {\bibinfo {volume} {73}},\ \bibinfo
  {pages} {085119} (\bibinfo {year} {2006})}\BibitemShut {NoStop}%
\bibitem [{\citenamefont {Barato}\ and\ \citenamefont
  {Seifert}(2015)}]{TUR-udo-first}%
  \BibitemOpen
  \bibfield  {author} {\bibinfo {author} {\bibfnamefont {A.~C.}\ \bibnamefont
  {Barato}}\ and\ \bibinfo {author} {\bibfnamefont {U.}~\bibnamefont
  {Seifert}},\ }\href {\doibase 10.1103/PhysRevLett.114.158101} {\bibfield
  {journal} {\bibinfo  {journal} {Phys. Rev. Lett.}\ }\textbf {\bibinfo
  {volume} {114}},\ \bibinfo {pages} {158101} (\bibinfo {year}
  {2015})}\BibitemShut {NoStop}%
\bibitem [{\citenamefont {Gingrich}\ \emph {et~al.}(2016)\citenamefont
  {Gingrich}, \citenamefont {Horowitz}, \citenamefont {Perunov},\ and\
  \citenamefont {England}}]{TUR-Gingrich}%
  \BibitemOpen
  \bibfield  {author} {\bibinfo {author} {\bibfnamefont {T.~R.}\ \bibnamefont
  {Gingrich}}, \bibinfo {author} {\bibfnamefont {J.~M.}\ \bibnamefont
  {Horowitz}}, \bibinfo {author} {\bibfnamefont {N.}~\bibnamefont {Perunov}}, \
  and\ \bibinfo {author} {\bibfnamefont {J.~L.}\ \bibnamefont {England}},\
  }\href {\doibase 10.1103/PhysRevLett.116.120601} {\bibfield  {journal}
  {\bibinfo  {journal} {Phys. Rev. Lett.}\ }\textbf {\bibinfo {volume} {116}},\
  \bibinfo {pages} {120601} (\bibinfo {year} {2016})}\BibitemShut {NoStop}%
\bibitem [{\citenamefont {Mohanta}\ \emph {et~al.}(2022)\citenamefont
  {Mohanta}, \citenamefont {Saryal},\ and\ \citenamefont
  {Agarwalla}}]{mohanta}%
  \BibitemOpen
  \bibfield  {author} {\bibinfo {author} {\bibfnamefont {S.}~\bibnamefont
  {Mohanta}}, \bibinfo {author} {\bibfnamefont {S.}~\bibnamefont {Saryal}}, \
  and\ \bibinfo {author} {\bibfnamefont {B.~K.}\ \bibnamefont {Agarwalla}},\
  }\href {\doibase 10.1103/PhysRevE.105.034127} {\bibfield  {journal} {\bibinfo
   {journal} {Phys. Rev. E}\ }\textbf {\bibinfo {volume} {105}},\ \bibinfo
  {pages} {034127} (\bibinfo {year} {2022})}\BibitemShut {NoStop}%
\end{thebibliography}%

\end{document}